\documentclass[twocolumn,preprintnumbers,superscriptaddress,amsmath,amssymb,prb]{revtex4-1}

\usepackage{graphicx}
\usepackage{dcolumn}
\usepackage{subfigure}
\usepackage{bm}
\usepackage[utf8]{inputenc}
\usepackage{comment}
\usepackage[unicode=true, bookmarks=false, breaklinks=false, pdfborder={0 0 1},colorlinks=false]{hyperref}
\usepackage{breakurl}
\usepackage{url}
\usepackage{natbib}
\usepackage{multirow}
\usepackage{float}

\usepackage{wrapfig}

\usepackage{xcolor}

\newcommand{\unit}[2][1]{#1~\mathrm{#2}}
\newcommand{\GRS}{GdRu$_2$Si$_2$}
\newcommand{\GRG}{GdRu$_2$Ge$_2$}

\begin{document}

\title{Magnetic frustration and non-collinear textures in layered Gd magnets}

\newcommand{\Uppsala}{Department of Physics and Astronomy, Uppsala University, Box 516, SE-75120 Uppsala, Sweden}
\newcommand{\KTH}{Department of Applied Physics, School of Engineering Sciences, KTH Royal Institute of Technology, AlbaNova University Center, SE-10691 Stockholm, Sweden}
\newcommand{\WISEUppsala}{Wallenberg Initiative Materials Science for Sustainability, Uppsala University,
75121 Uppsala, Sweden}
\newcommand{\WISEKTH}{Wallenberg Initiative Materials Science for Sustainability (WISE), KTH Royal Institute of Technology, SE-10044 Stockholm, Sweden}
\newcommand{\SeRC}{SeRC (Swedish e-Science Research Center), KTH Royal Institute of Technology, SE-10044 Stockholm, Sweden}

\author{Vladislav Borisov}
    \affiliation{\Uppsala}
    \affiliation{\WISEUppsala}
    \email[Corresponding author:\ ]{vladislav.borisov@physics.uu.se}

\author{Rohit Pathak}
    \affiliation{\Uppsala}
    \affiliation{\WISEUppsala}

\author{Sagar Sarkar}
    \affiliation{\Uppsala}

\author{Anna Delin}
    \affiliation{\KTH}
    \affiliation{\WISEKTH}
    \affiliation{\SeRC}

\author{Olle Eriksson}
    \affiliation{\Uppsala}
    \affiliation{\WISEUppsala}
    
\date{\today}

\begin{abstract}
Using scale-bridging simulations based on electronic structure theory and atomistic spin dynamics, we investigate the magnetic properties of skyrmionic GdRu$_2$Si$_2$ and GdRu$_2$Ge$_2$ layered rare-earth magnets and similar Gd-based compounds (GdAu$_2$Si$_2$, GdAu$_2$Ge$_2$, GdAg$_2$Si$_2$ and GdAg$_2$Ge$_2$). By studying the trends across this structural family, we confirm the importance of magnetic frustration and dipolar interactions for the stability of non-collinear and skyrmion phases. Furthermore, our calculations predict promising opportunities for chemical tuning of these magnets in terms of the balance between various exchange interactions and the character of the magnetic anisotropy. These changes lead to the formation of new types of skyrmions that are stable in a wide range of applied external magnetic field. In particular, we propose partial alkali-metal substitution of Gd in GdRu$_2$Si$_2$ leading to GdKRu$_4$Si$_4$, GdRbRu$_4$Si$_4$, GdCsRu$_4$Si$_4$ as well as GdYRu$_4$Si$_4$ compounds, and suggest that it is likely to result in an ordered layered structure, similarly to previously reported iron pnictides like CaKFe$_4$As$_4$.
\end{abstract}

\maketitle

\section{Introduction}

\textbf{Literature overview.} A flurry of research activity was prompted by the discovery of magnetic skyrmions in a solid-state system MnSi with the chiral B20 structure, reported in a seminal work in 2009 [\onlinecite{Muehlbauer2009}]. The large interest in skyrmions is tightly connected to their real-space topology of the spin configuration, which brings about important advantages for future spintronics applications, such as memory storage and faster energy-efficient computations [\onlinecite{Fert2013,Huang2017,Bourianoff2018,Pinna2018}].

Usually, magnetic skyrmions are stabilized by the Dzyaloshinskii-Moriya (DM) interaction which favors a certain chirality (left- or right-handed) of the spin texture. There are, however, examples where exceptionally small skyrmions $\sim\!\unit[2]{nm}$ stabilize due to the interplay of frustrated Heisenberg interactions and on-site anisotropy, while the DM interaction is strictly zero due to the presence of structural inversion symmetry. One such system is \GRS{} [\onlinecite{Khanh2020}] which has a layered structure of 122-type (also known as the widespread ThCr$_2$Si$_2$ structure type), similar to the well-known structural family of iron pnictides such as the collinear antiferromagnet CaFe$_2$As$_2$ [\onlinecite{Torikachvili2008}]. The difference is that \GRS{} shows a wider variety of magnetic phases, including skyrmions, formed by large Gd moments, which interact through the conduction electrons.

In the literature, frustrated Heisenberg interactions in \GRS{} are often associated with the RKKY mechanism [\onlinecite{Ruderman1954},\onlinecite{Kasuya1956},\onlinecite{Yosida1957}]. Since the RKKY exchange interaction oscillates strongly in space, the magnitude and even the sign of magnetic interactions in \GRS{} can be very sensitive to structural details, such as the distance between Gd cations in each layer, the interlayer separation and the Si crystallographic position. This is due to the dependence of the RKKY interaction on the shape of the Fermi surface that depends on the crystal structure. This is clearly shown in our work [\onlinecite{Sarkar2026}] where the effects of uniaxial pressure on the magnetic phase diagram of GdRu$_2$Si$_2$ are modelled and also discussed in [\onlinecite{Yoshimochi2024}] where the role of RKKY interaction is considered for the emergence of skyrmion spin textures.

On the other hand, crystal structure can be tuned not only by mechanical pressure but also by chemical doping, and it was considered in literature [\onlinecite{Nomoto2023}] to replace the Gd, Ru and Si species by different elements, e.g.~Ru by Ag or Si by Ge as well as to consider Eu-based 122 compounds. Many such systems were analyzed theoretically in that paper from the point of view of Heisenberg exchange interactions in real space ($J_{ij}$ interaction between spins $i$ and $j$) and Fourier space (spin-spiral energy $E(q)$ for the whole system). Several candidate compounds, such as GdAg$_2$Si$_2$ and GdAg$_2$Ge$_2$, were proposed as new systems with spiral magnetic order that can potentially be transformed into skyrmions. This statement, however, requires a theoretical confirmation, since the corresponding simulations of equilibrium magnetic configuration were not yet performed in [\onlinecite{Nomoto2023}].

\textbf{Goals of this work.} In the present work, we study further the known compounds like \GRS{} and Ag/Au-based systems with the same type of structure but also suggest new layered rare-earth magnets with the so-called 1144-type structure, similar to superconductors like CaKFe$_4$As$_4$. Such compounds where the rare-earth magnetic layers are alternating not just with Ru$_2$Si$_2$ but also with non-magnetic alkali metal layers (\textit{A} = K, Rb or Cs) have so far not been considered as skyrmionic systems in the literature. However, the possibility of ordered alternating-layer structures is strongly supported by previous studies of pnictide superconductors such as CaKFe$_4$As$_4$, where structural stability depends on the large ionic size difference between Ca and K cations. One motivation for forming 1144-type compounds Gd\textit{A}Ru$_4$Si$_4$ is to increase the interlayer distance between Gd planes and make the system more two-dimensional, which can change the magnetic properties. Furthermore, alkali metal doping will change the electron count and is expected to influence the microscopic magnetic interactions and larger-scale magnetic properties. To partially decouple the structural effects (interlayer separation) from the electronic effects (carrier doping), we also consider the GdYRu$_4$Si$_4$ compound which, due to nominal 3+ charge of Y, has the same electron count as the GdRu$_2$Si$_2$ system.

For all these systems, we study the magnetic phase diagrams based on atomistic spin-dynamics simulations (ASD [\onlinecite{uppasd},\onlinecite{Eriksson2017}]) with magnetic interactions calculated using magnetic force theorem and density functional theory (further details in section~II). Such an approach allows to address magnetic phenomena on different length scales fully from first principles (see e.g.~[\onlinecite{Szilva2023},\onlinecite{Borisov2024rev}]). The obtained numerical results give insights into which of these systems can host interesting non-collinear or even topological magnetic phases, which is the main point of this work.

\section{Theory and simulation methods}

Our final goal, i.e.~to predict the magnetic phase diagrams of the studied systems, is achieved in three steps in this work: the first step focuses on  electronic structure calculations, and the second step on computations of the magnetic interaction parameters. Finally, in the third step we perform atomistic spin-dynamics simulations at varying external field.

First, we determine the electronic structure from first principles using density functional theory [\onlinecite{Hohenberg1964}] available in the Full-potential Linear Muffin-Tin Orbital RSPt software [\onlinecite{Wills1987},\onlinecite{Wills2010}]. The 4$f$ states of Gd are described in the frozen-core approximation, which works quite well for that element as discussed in literature [\onlinecite{Locht2016}]. The calculated electronic properties allow, for example, to evaluate the ionic forces, which is necessary for optimizing the crystal structure. The structure optimization is done here using the projector-augmented wave method of VASP code for all compounds, except for GdRu$_2$Si$_2$ and GdRu$_2$Ge$_2$ where also experimental structures are available [\onlinecite{kresse1996efficient}] (further details are in Appendix~B). Electronic properties are also used in the next step where the Heisenberg and DM magnetic interactions between Gd spins are calculated using the LKAG approach (magnetic force theorem) [\onlinecite{LKAG1987}]. Also, the on-site anisotropy is evaluated based on eigenvalue sums for different orientations of the spin axis (along the $x$-, $y$- and $z$-directions).

The resulting effective spin model, which describes approximately the magnetic energy of the systems, can be summarized as follows:
\begin{align}
    H = &-\sum\limits_{\langle i, j \rangle} \left[ J_{ij} (\vec{e}_i \cdot \vec{e}_j) + \vec{D}_{ij} \cdot (\vec{e}_i \times \vec{e}_j) \right] +\\ 
    & +K_U\! \sum\limits_i (\vec{e}_i \cdot \hat{z} )^2 -g\mu_B m \sum\limits_i (\vec{e}_i \cdot \vec{B}_\mathrm{ext}),
    \label{e:spin_model}
\end{align}

where $\vec{e}_i$ are unit vectors describing the orientation of magnetic moments of individual atomic sites and the summation $\langle i,j \rangle$ runs over unordered pairs of spins (which implies that $i \neq j$); $m$ is the magnitude of Gd moments. Regarding the exchange parameters, the convention here is such that positive $J_{ij}$ corresponds to a ferromagnetic coupling and negative $K_U$ corresponds in uniaxial anisotropy that prefers spins to lie along the $\vec{z}$-direction, i.e.~perpendicular to the Gd layers. For the systems that will be discussed in the following, the parameters of this Hamiltonian (summarized for nearest neighbors in Tables~I and~II in Appendix~A) have varying signs depending on the distance in the crystal lattice and chemical composition, suggesting good possibilities to observe a large variety of magnetic phases.

As will be demonstrated in the results section, the dipolar interactions are important in some of the studied systems and change significantly the character of magnetic states. To take into account these interactions, we include the following well-known term in the Hamiltonian:
\begin{equation}
H_{\mathrm{dip}}
=
\frac{\mu_0}{4\pi}
\sum_{\langle i,j\rangle}
\frac{m^2}{R_{ij}^{3}}
\left[
\mathbf{e}_i\cdot\mathbf{e}_j
-
3\left(\mathbf{e}_i\cdot\hat{\mathbf{r}}_{ij}\right)
 \left(\mathbf{e}_j\cdot\hat{\mathbf{r}}_{ij}\right)
\right],
\label{e:dipolar_exchange}
\end{equation}

Here, $\mu_0$ is the vacuum magnetic permeability and $\hat{r}_{ij}$ is the unit vector connecting two Gd spins with magnetic moments $m$ along the unit vectors $\vec{e}_i$ and $\vec{e}_j$, while $R_{ij}$ is the distance between those spins. Additional prefactor $\frac{1}{2}$ is included in Eqn.~(\ref{e:dipolar_exchange}) as well to account for the double counting in the summation over different spins. To calculate the dipolar interactions as accurately as possible without making the calculations computationally unfeasible, we used here the FFT-based implementation available in the UppASD code. In this approach, described in Ref.~\onlinecite{Hayashi1996}, the real-space dipolar interaction coefficients and spin distribution on the atomic sites of the supercell are Fourier-transformed and then the convolution theorem is used to do the summation over all sites. The computational complexity scales here better compared to direct real-space summation. Furthermore, Ewald summation is part of this FFT-implementation meaning that the long-range part of dipolar interactions is included to a good accuracy.

While dipolar interactions are basically negligible for strong permanent magnets (except for their domain structure), they can contribute visibly to the magnetic properties of compounds with weak Heisenberg interactions and small anisotropy, which is the case for GdRu$_2$Si$_2$ and its derivatives. One can especially notice that Gd bears large moments around $\unit[7]{\mu_\mathrm{B}}$ which are placed in a layered structure. As our model calculations based on eqn.~(\ref{e:dipolar_exchange}) show, the asymmetry between the intralayer and interlayer Gd-Gd distances due to non-zero tetragonality leads to enhanced contribution of dipolar interactions to the total energy (further discussion is in Section~III).

In the final step, the spin model (\ref{e:spin_model}) together with dipolar interactions (\ref{e:dipolar_exchange}) is inserted in the Landau-Lifshitz-Gilbert equation [\onlinecite{Landau1935},\onlinecite{Gilbert2004}] to simulate the atomistic spin dynamics (ASD) using the UppASD software [\onlinecite{uppasd},\onlinecite{Eriksson2017}]:
\begin{equation}
    \frac{\partial \vec{m}_i}{\partial t} = -\frac{\gamma}{1 + \alpha^2} \left( \vec{m}_i \times \vec{B}_i + \frac{\alpha}{m}\,\vec{m}_i \times (\vec{m}_i \times \vec{B}_i) \right)
    \label{e:LLG_equation}
\end{equation}

The magnetic moments of individual sites ($\vec{m}_i$) are defined here in the same way as in Eqn.~(\ref{e:dipolar_exchange}); $\alpha$ is the Gilbert damping parameter which is a free parameter in the annealing and measurement phases of the ASD simulation. Importantly, for the here studied systems we find that relatively large time steps up to tens of picoseconds still provide numerically reliable ASD results with respect to the final magnetic state at zero temperature. We set this time step to 1\,ps in the annealing procedure and to 2\,ps in the measurement phase at $T=0$. The effective magnetic field $\vec{B}_i$ in Eqn.~(\ref{e:LLG_equation}) is obtained from the spin model (\ref{e:spin_model}) as a minus derivative of the sum of expressions (\ref{e:spin_model}) and (\ref{e:dipolar_exchange}) (see e.g.~book [\onlinecite{Eriksson2017}] and review [\onlinecite{Borisov2024rev}] for details).

Also, the supercell dimensions used in the ASD simulations matter, since they should be large enough to accommodate different magnetic states but not too large to keep the simulations computationally feasible. For GdRu$_2$Si$_2$ we found optimal supercell to have $(60\times 60\times 10)$ dimensions, relative to the 2-Gd unit cell, while for GdRu$_2$Ge$_2$ it was sufficient to work with the $(40\times 40\times 10)$ supercell. For derivative systems, like GdRu$_2$GeSi and 1144-type compounds, we use $(50\times 50\times 10)$ supercells.

It should be emphasized that throughout this work our simulations are done assuming non-periodic boundary conditions, which gives a flexibility to accommodate different types of magnetic states. The final equilibrated states often show a boundary region with a thickness of a few unit cells where the magnetization pattern is very different due to the surface effects (e.g.~reduced number of neighbors). However, the inside of the simulation supercell reveals well-defined patterns representing the bulk systems.
\begin{figure*}
\vspace{10pt}
{\centering
\includegraphics[width=0.95\textwidth]{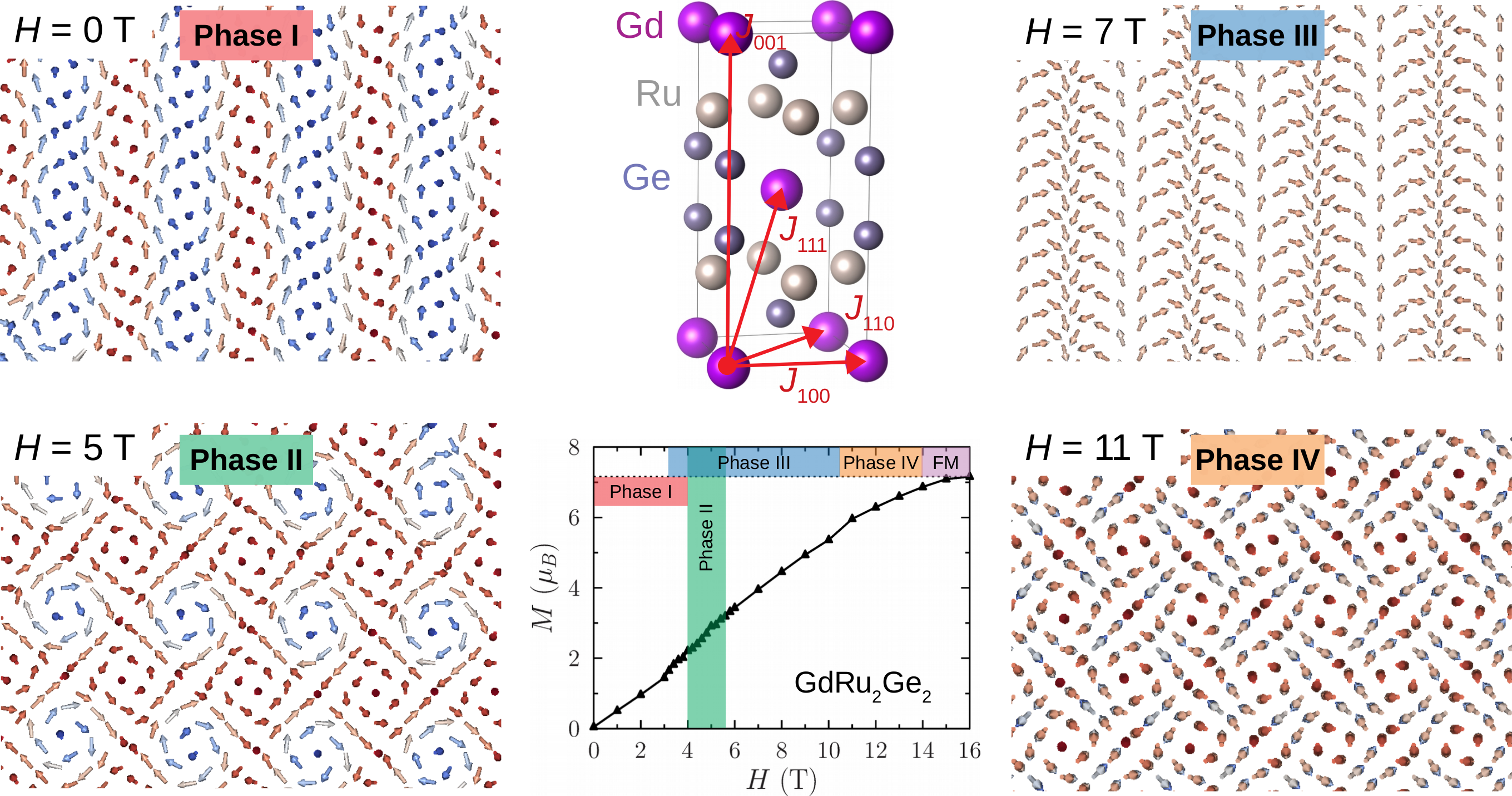}
}
\caption{Variety of magnetic phases in GdRu$_2$Ge$_2$ found in numerical atomistic spin dynamics simulations based on the calculated effective spin model. The magnetic phase diagram in the bottom center shows the total magnetization vs applied field together with existence regions for different magnetic configurations (phases I--IV which include spin spirals and skyrmion textures) as well as the ferromagnetic state. The snapshots of phases I--IV (spin configurations of individual Gd layers) are shown in the respective plots on the left and right sides.}
\label{f:GdRu2Ge2}
\end{figure*}

To augment the analysis of equilibrium spin configurations, the topological charge is calculated for each square plaquette of nearest-neighbor Gd magnetic moments according to the definition suggested in [\onlinecite{Kim2020}]:
\begin{equation}
    Q_\mathrm{local} = \frac{1}{4\pi} ( q_{124} + q_{234} ),
    \label{e:charge}
\end{equation}
where the quantity $q_{ijk}$ is defined for each of the two triangles (1-2-4 and 2-3-4), that make up the considered plaquette of 4 Gd spins (see figure~2 in [\onlinecite{Kim2020}]), as follows:
\begin{equation}
    \tan\!\left(\ \!\!\frac{q_{ijk}}{2} \right) = \frac{ \vec{e}_i\cdot (\vec{e}_j\times \vec{e}_k) }{ 1 + \vec{e}_i\cdot \vec{e}_j + \vec{e}_i \cdot \vec{e}_k + \vec{e}_j \cdot \vec{e}_k }
\end{equation}

The local topological charge defined by Eqn.~(\ref{e:charge}) is plotted then as a 2D map for each plaquette in one of the Gd layers in the simulation cell (see Figs.~\ref{f:topological_charge_GdRu2Si2}--\ref{f:topological_charge_1144} in Appendix~C). Comparing this real-space distribution of topological charge with the spin configuration helps to draw conclusions about the topological character of various magnetic phases that are predicted for the studied rare-earth systems.

\section{$\mathrm{Ru}$-based 122-type compounds}

\textbf{Magnetic phase diagram.} For the example of GdRu$_2$Ge$_2$ with experimental crystal structure (also sketched in the middle of Fig.~\ref{f:GdRu2Ge2}) we discuss the variety of magnetic states that are obtained in our simulations at zero and applied magnetic field:

1) \textbf{Phase I} at low field ($H \leq \unit[4]{T}$) is a \textit{spin spiral} that coexists with phases II and III (see the following points). This spin spiral is corrugated, since it is not a purely single-$\vec{Q}$ state but has a finite contribution from another $\vec{Q}$-vectors. We get the same phase~I for GdRu$_2$Si$_2$, which is in line with recently reported experimental observations [\onlinecite{Wood2023},\onlinecite{Spethmann2024}]. However, experimental data for GdRu$_2$Ge$_2$ [\onlinecite{Yoshimochi2024}] indicates that phase~I in that compound is a single-$\vec{Q}$ magnetic state, which does not coexist with other phases. This comparison of two systems, both on theory and on experimental levels, presents an interesting case for a future study.

2) \textbf{Phase II} (\textit{square skyrmion lattice}) appears between $\unit[4.0-5.6]{T}$ but most skyrmions are observed between $\unit[4.6-5.0]{T}$, so the stability region for this magnetic phase is quite narrow compared to other phases, in line with the first experiment [\onlinecite{Khanh2020}]. The topological number here is non-zero and it can have different signs, because DM interaction is absent in the system and for that reason both types of magnetic chirality have the same energy. Fig.~\ref{f:GdRu2Ge2} shows a closeup of this phase at $H = \unit[5]{T}$.

3) \textbf{Phase III} is a different \textit{spin spiral} with a finite average magnetization along the field direction (perpendicular to the image). This phase starts appearing at $H = \unit[3.2]{T}$ and completely vanishes after $\unit[10]{T}$.

4) At higher fields $\unit[11]{T} \leq H \leq \unit[15]{T}$, we only see \textbf{Phase IV} which has a large $m_z$ component, while the $m_x$ and $m_y$ components follow a square skyrmion texture reminiscent of phase~II. Due to the large non-zero average $m_z$ component, this magnetic phase is \textit{topologically trivial}, since the spin vectors do not wrap around the whole unit sphere (see detailed discussion of this aspect in the literature [\onlinecite{Braun2012},\onlinecite{Kim2020}]), but the real-space-resolved topological charge (Fig.~\ref{f:topological_charge_GdRu2Ge2}g) indicates a double-$\vec{Q}$ character of this magnetic state. The high-field phase IV is suppressed above $\unit[15]{T}$, as it turns into the ferromagnetic (FM) phase where $m_z$ is saturated.

\begin{figure}
\vspace{10pt}
{\centering
\includegraphics[width=0.42\textwidth]{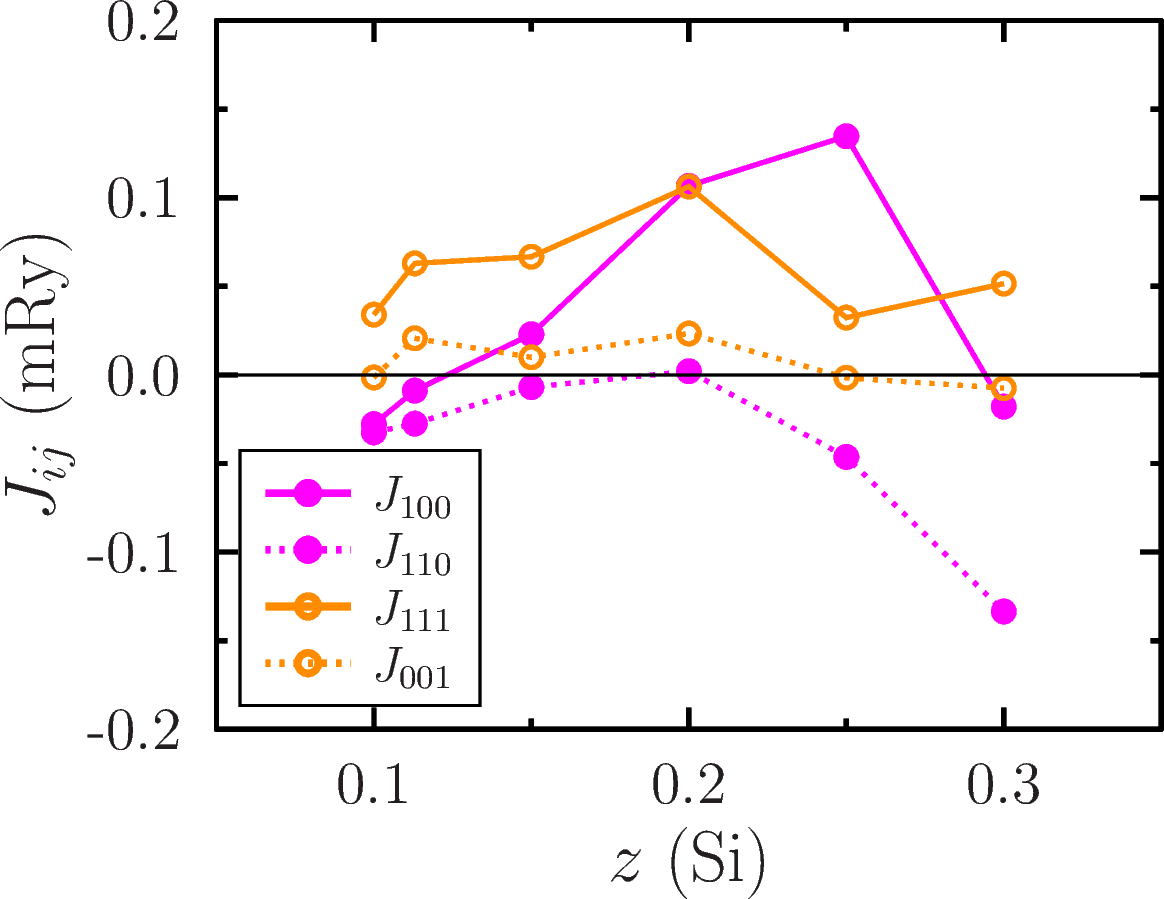}
}
\vspace{-7pt}
\caption{Nearest-neighbor magnetic Heisenberg interactions in GdRu$_2$Si$_2$ calculated for different $z$-coordinates of the Si sites. $J_{ij} > 0 $ corresponds to ferromagnetic interaction and $J_{ij} < 0 $~-- to antiferromagnetic one.}
\label{f:Jij_vs_zSi}
\end{figure}
All these phases, except for the FM one, are sketched in Fig.~\ref{f:GdRu2Ge2} where also the stability regions for phases I--IV are marked in the central $M-H$ plot of total magnetization vs applied magnetic field.

\begin{figure*}
\vspace{10pt}
{\centering
\includegraphics[width=0.95\textwidth]{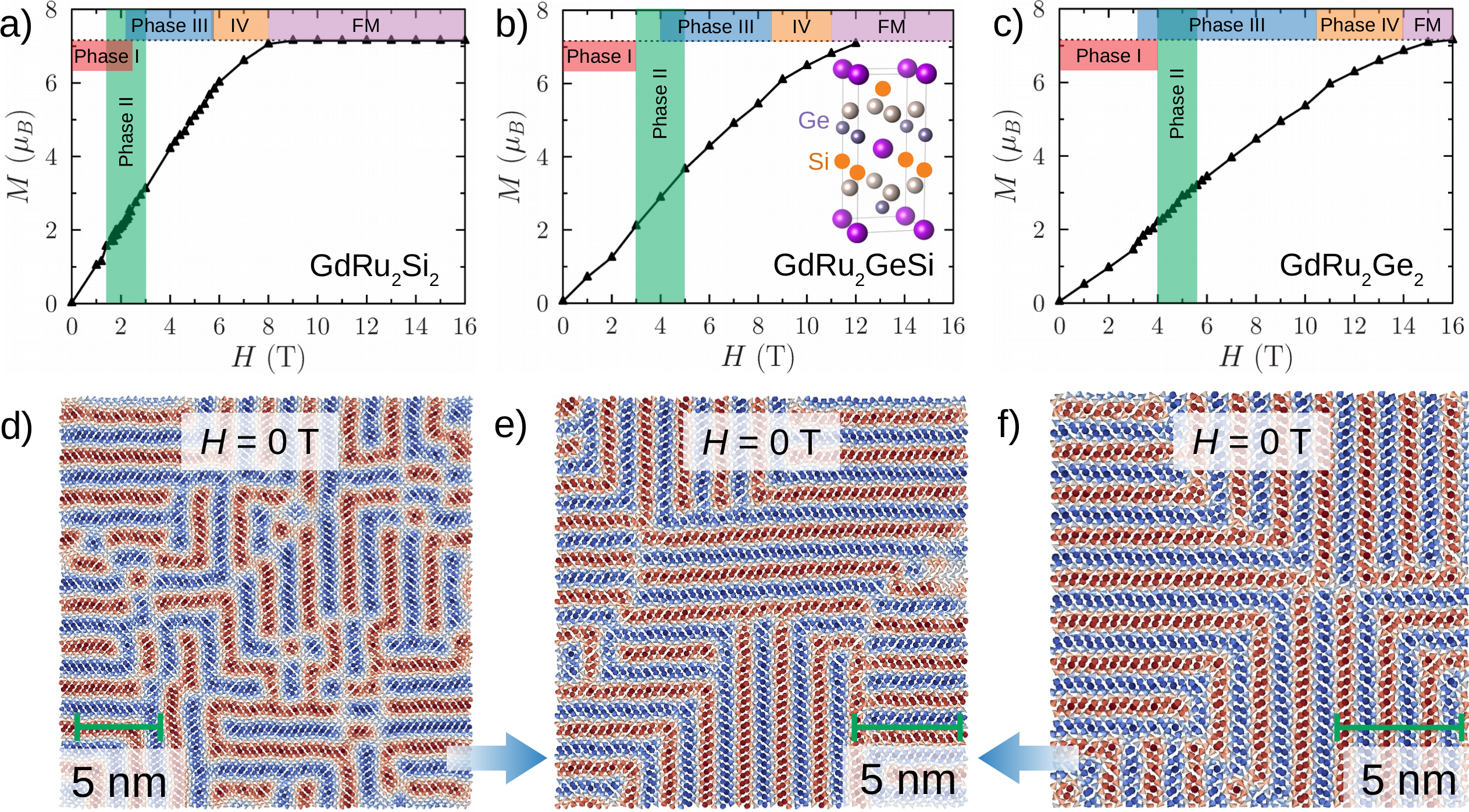}
}
\caption{Magnetic trends in the GdRu$_2$Si$_2$--GdRu$_2$GeSi--GdRu$_2$Ge$_2$ series, based on atomistic spin dynamics results. The top plots represent the magnetization-vs-field dependence for each system (and crystal structure for GdRu$_2$GeSi specifically), while the bottom plots show the equilibrium magnetic configurations in individual Gd layers at zero field. Note the slightly different length scales in the bottom plots, indicated by the green 5-nm bars.}
\label{f:GdRu2GeSi}
\end{figure*}
\textbf{Model robustness.} Important remarks should be made here about the calculated magnetic interactions and parameters that enter the spin model (\ref{e:spin_model}). These first-principles model parameters are very sensitive to the structural details, such as the Si position, lattice parameters which directly affect the Gd-Gd distances and the Fermi surface structure. The later is discussed in the literature [\onlinecite{Bouaziz2022}] to be crucial for such compounds due to the RKKY character of Heisenberg magnetic interactions. In particular, our test calculations reveal that the Gd-Gd interactions can even change sign due to small variations in any of the aforementioned structural parameters. For example, the nearest-neighbor Gd-Gd exchange in each Gd layer of GdRu$_2$Ge$_2$ becomes ferromagnetic instead of antiferromagnetic when the crystal structure is optimized using the GGA functional. In that case, e.g.~phase~II could not be stabilized in spin-dynamics simulations. Using the PBEsol functional, we get a slightly better agreement with experiment for the $c$-lattice parameter but worse for the $a$-lattice parameter, while the Heisenberg Gd-Gd interactions calculated within GGA for this PBEsol structure are still similar (nearest-neighbor FM, next-nearest-neighbor AFM). The magnetic interactions do not change much also for the GGA-optimized structure where a negative pressure of $\unit[3]{GPa}$ is imposed, which allows to match the $a$-lattice parameter with experiment at the expense of increasing the $c$-lattice parameter.

Another series of tests that we performed is concerning the GdRu$_2$Si$_2$ compounds where we fix the lattice parameters and vary the position of the Si ions. In Fig.~\ref{f:Jij_vs_zSi}, one can see that the nearest-neighbor Heisenberg interactions, along the [100], [110], [111] and [001] directions, can even change the sign depending on the Si position, which leads to a totally different magnetic behavior than what is observed experimentally. For that reason, one can expect strong spin-lattice coupling effects in this kind of systems and, on the other hand, significant theory challenges in terms of describing accurately their crystal and magnetic structures, for example, under mechanical or chemical pressure. Discussion of this aspect will be continued in section~V. Based on our simulations we can say that the experimental crystal structures of GdRu$_2$Si$_2$ and GdRu$_2$Ge$_2$ give the best set of magnetic parameters for further simulations in terms of reproducing different magnetic phases (spirals, skyrmions~etc.) that mostly agree with experimental findings from the literature [\onlinecite{Khanh2020},\onlinecite{Yoshimochi2024}].

\textbf{Importance of dipolar interactions and local anisotropy.} By comparing the spin dynamics results at different values of external field with and without dipolar interactions we see that these dipolar interactions are absolutely crucial for stabilizing the skyrmion lattice (phase II) and skyrmion-like high-field phase IV. Without dipolar interactions it is not possible to stabilize these two phases, which are reported in experiments [\onlinecite{Yasui2020,Wood2023,Spethmann2024}]. Also the corrugated double-$\vec{Q}$ structure of phase~I relies on the effect of dipolar interactions. Regarding the on-site magnetic anisotropy (last term in Eqn.~\ref{e:spin_model}), based on test calculations, where the dipolar exchange is included, we found that it does not affect the calculated magnetic ground state for the range of values between $0$ and the first-principles DFT value for the experimental crystal structure of GdRu$_2$Ge$_2$.

\textbf{Annealing procedure.} Furthermore, to obtain clearly shaped and well-ordered extended domains of square skyrmion lattices, it is important to do a slow annealing procedure with several temperature steps and many time iterations in the spin dynamics. The other magnetic phases are more robust in this respect and can be stabilized even with faster annealing, but the skyrmion lattice phase in these systems is really sensitive to the annealing protocol. This is also in line with the experimental observation that the skyrmion phase occupies only a small part of the magnetic phase diagram (see Fig.~1b in [\onlinecite{Khanh2020}]), most likely due to the fine balance between different interactions that stabilize the skyrmion lattice.

All the statements we made in this section hold also for GdRu$_2$Si$_2$, which is qualitatively similar to GdRu$_2$Ge$_2$ (see e.g.~Fig.~\ref{f:GdRu2GeSi}a,c and topological charge in Fig.~\ref{f:topological_charge_GdRu2Si2}).

\vspace{5pt}
\textbf{Si-Ge substitution.} While the two 122-type compounds discussed above look similar from the magnetic point of view, they show some important quantitative differences. For example, GdRu$_2$Si$_2$ shows uniaxial anisotropy $\sim\!\unit[0.05]{meV/Gd}$, while the anisotropy is an order of magnitude smaller in GdRu$_2$Ge$_2$. Also, the calculated saturation field for GdRu$_2$Si$_2$ is factor of 2 lower than for GdRu$_2$Ge$_2$ (see Fig.~\ref{f:GdRu2GeSi}a,c), and the field range for skyrmion stability is shifted by the same factor.

In view of this, it is interesting to look at trends in the magnetic properties when substituting Si with Ge. As an example, we considered a hypothetical GdRu$_2$GeSi compound with alternating Ru$_2$Ge$_2$ and Ru$_2$Si$_2$ blocks (see inset in Fig.~\ref{f:GdRu2GeSi}b). This example structure is an ordered approximation of what one could expect an experimental GdRu$_2$GeSi compound should look like (which is not synthesized yet), with a randomness of the occupation of Ge and Si atoms. However, as we will see in the following the ordered structure helps to reveal interesting trends.

First, we notice that the structural parameters ($a$ and $c$ lattice constants and $X$-$X$ interlayer distance ($X$ = Si, Ge) for GdRu$_2$GeSi are in-between the corresponding values for the pure compounds. This is as far as the theoretically obtained structural parameters are concerned. The theory values, however, are overestimated compared to the measured ones, which is a well-known underbinding problem of density functional theory with the GGA functional. Especially for the interlayer Si-Si and Ge-Ge distances the mismatch between theory and experiment is large, and, as we discussed in section~III, this affects critically the calculated magnetic interactions. In this respect, the trends in the \GRS{}--GdRu$_2$GeSi--\GRG{} series may be different depending on whether the structures of individual compounds are optimized by theory or taken from measurements.

For this reason, we have performed calculations for a hypothetical structure of GdRu$_2$GeSi which is obtained by linear interpolation between the measured structures of \GRS{} and \GRG{}. The interpolation is done separately for the $a$ and $c$ lattice parameters and Si-Si distance $d$ across the Gd layers based on the following equations:
\begin{align*}
    a(\mathrm{GeSi}) &= \alpha_a\cdot a(\mathrm{Ge}) + (1 - \alpha_a)\cdot a(\mathrm{Si}) \\
    c(\mathrm{GeSi}) &= \alpha_c\cdot c(\mathrm{Ge}) + (1 - \alpha_c)\cdot c(\mathrm{Si}) \\
    d(\mathrm{GeSi}) &= \alpha_d\cdot d(\mathrm{Ge}) + (1 - \alpha_d)\cdot d(\mathrm{Si})
\end{align*}

Here, the interpolation parameter $\alpha$ is different for $a$, $c$ and $d$ and is calculated from these equations applied to the theoretically optimized structures. This gives us the values $\alpha_a = 0.294$, $\alpha_c = 0.692$ and $\alpha_d = 0.377$. Afterwards, these $\alpha$ values are used together with experimental $a$, $c$ and $d$ values for \GRS{} and \GRG{} to obtain an interpolated structure of GdRu$_2$GeSi.

Based on the calculated magnetic parameters of the mixed Si-Ge and pure Si- and Ge-systems, we have first calculated the adiabatic magnon spectra (Fig.~\ref{f:magnons} in Appendix~B) which reveal similarity between all 3 systems in terms of magnon energy dispersions and presence of imaginary frequencies between $\Gamma-X$ points. These indicate magnetic instability towards a helical magnetic configuration, which is, indeed, close to the actual ground states of \GRS{} and \GRG{}. In terms of magnons, the mixed Si-Ge system is more similar to the pure Ge system.

In the next step, we use the same magnetic interaction parameters to perform atomistic spin-dynamics simulations. These have produced similar zero-field ground states dominated by helical spin spiral with a nanometer length scale. A skyrmion phase also appears in our simulations for the GdRu$_2$GeSi compound within a similar field range between $\unit[(2-5)]{T}$. The magnetization versus applied field plot is qualitatively the same as well, only that the saturation field is around $\unit[12]{T}$, right in-between the corresponding values for the GdRu$_2$Si$_2$ and GdRu$_2$Ge$_2$ compounds. All this suggests that the variation of magnetic properties due to Ge substitution of Si is rather smooth and monotonous, most likely without any qualitative changes.

\section{$\mathrm{Ag}/\mathrm{Au}$-based 122 systems}

\begin{figure}
{\centering
\includegraphics[width=0.45\textwidth]{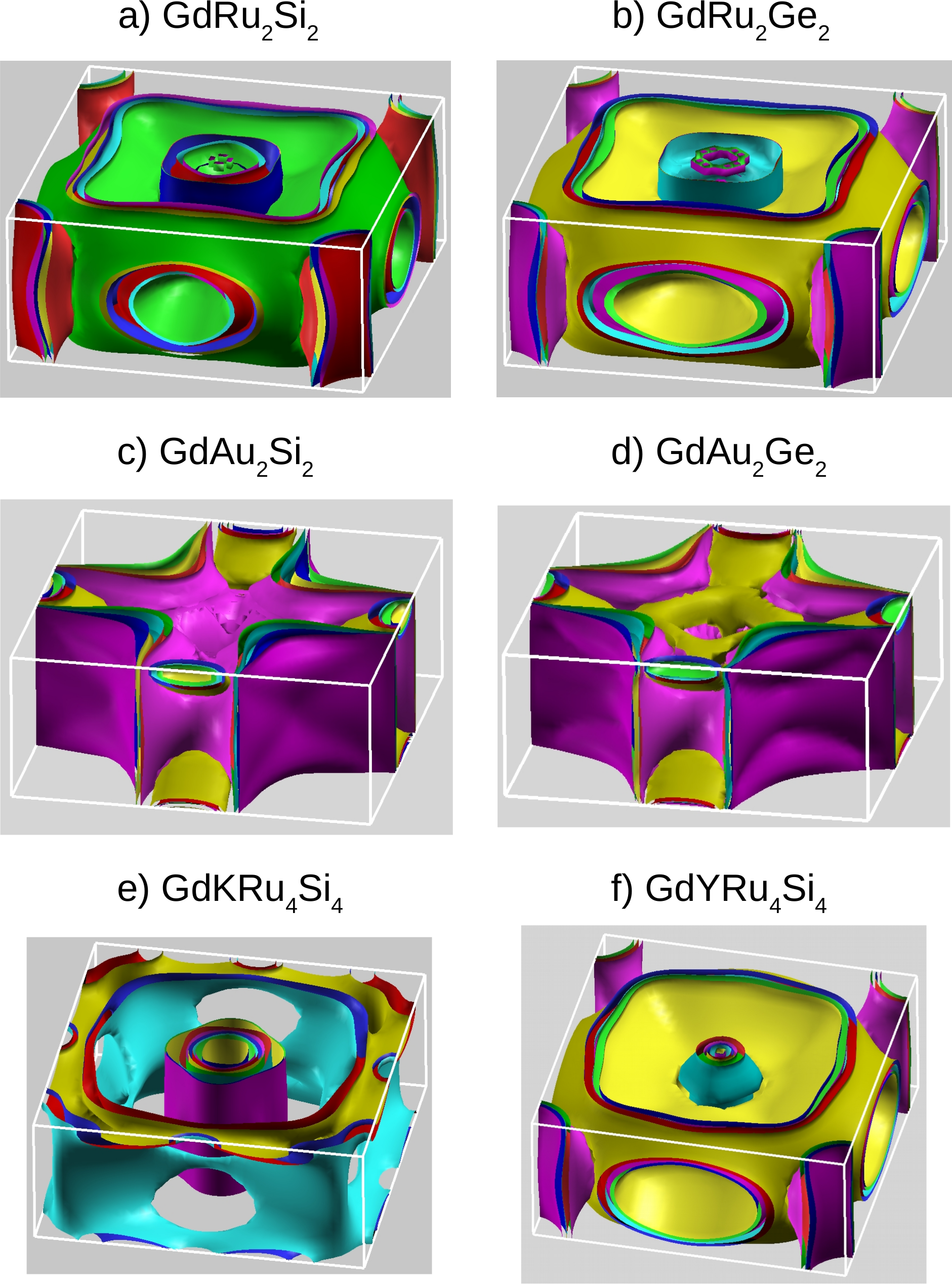}
}
\caption{Fermi surfaces (FS) of different 122- and 1144-type compounds calculated using density functional theory. The GdKRu$_4$Si$_4$ FS is visibly more 2D compared to the 122-type compounds, while GdYRu$_4$Si$_4$ shows similarity to the later due to the matching electron count.}
\vspace{-10pt}
\label{f:Fermi_surfaces}
\end{figure}
Similarly to the GdRu$_2$\textit{X}$_2$ compounds (\textit{X} = Si, Ge) that we analyzed in Section~III, we have also investigated the effect of substituting Ru with Ag or Au, leading to a subfamily of compounds: GdAg$_2$Si$_2$, GdAg$_2$Ge$_2$, GdAu$_2$Si$_2$ and GdAu$_2$Ge$_2$. In Ref.~\onlinecite{Nomoto2023}, these compounds were proposed to be possible candidates for nanoscale skyrmions, similarly to \GRS{}, and this statement was based on spin-spiral calculations which indicated a tendency to non-collinear spin-spiral ordering. However, the actual magnetic ground state in applied field was not studied, and so far they are not available experimentally. To verify this proposal, we have calculated in detail the Heisenberg spin Hamiltonian of these 4 systems, using theoretically optimized structures (see methodology in Appendix~B), and performed spin dynamics simulations to see what the magnetic ground state of these systems looks like and how it evolves with applied magnetic field, again compared to the well-known \GRS{} and \GRG{} skyrmionic compounds.

\begin{figure}
{\centering
\includegraphics[width=0.45
\textwidth]{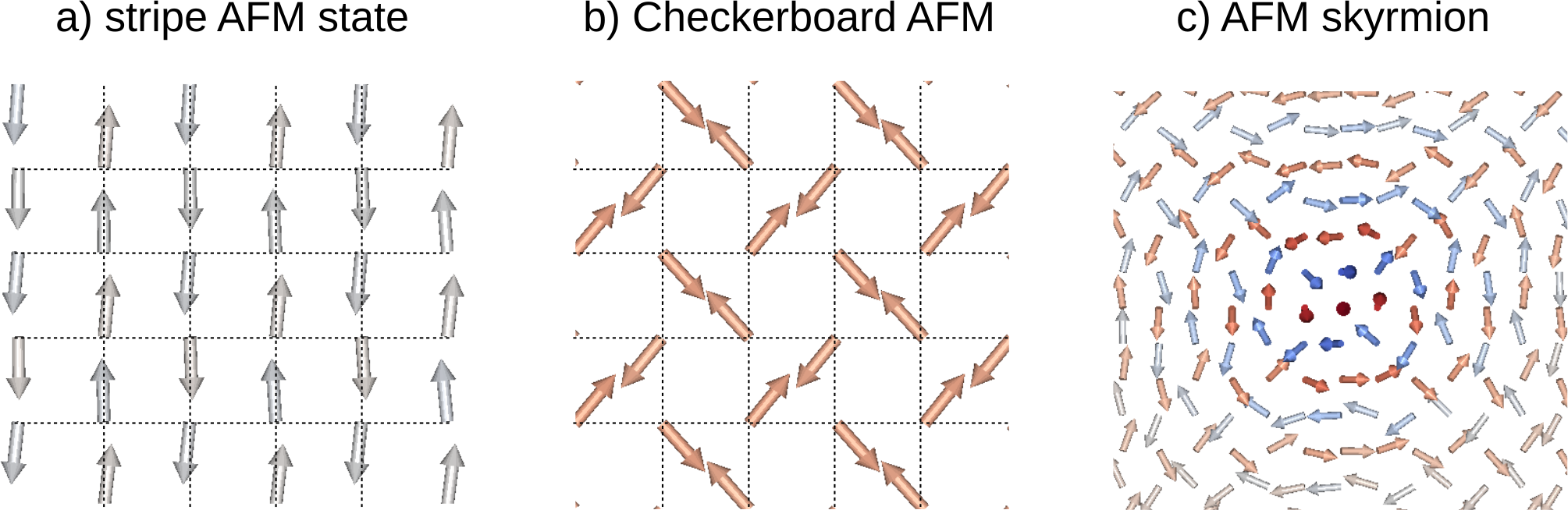}
}
\vspace{-5pt}
\caption{Snapshots of simulated zero-temperature spin configurations in individual Gd layers for the Au/Ag-based 122-type systems at different applied magnetic fields. Shown here are selected parts of simulation cells that may contain a few of these spin configurations simultaneously as domains or impurities (in case of skyrmion). Spin configurations ``a'' and ``b'' are shown by all compounds, except for GdAg$_2$Ge$_2$ which hosts skyrmion-like textures at all fields before becoming ferromagnetic. These skyrmion objects ``c'' can be stabilized also in GdAu$_2$Si$_2$ and GdAu$_2$Ge$_2$ at certain applied fields (see details in the text).}
\vspace{-15pt}
\label{f:Au_Ag_systems}
\end{figure}

It should be noted that these 4 compounds contain 3 more electrons per formula unit compared to \GRS{} due to Ag/Au substitution, which has a direct effect on the Fermi surface (examples are shown in Fig.~\ref{f:Fermi_surfaces}c,d). The later turns out to be qualitatively different from that of \GRS{} and other compounds studied in this work. Also, in case of Au-based systems, the $5d$ electrons are expected to be more delocalized than the $4d$ electrons in Ag- and Ru-based compounds.

From the calculated Heisenberg exchange interactions (Table~I),  we deduce that the 4 systems are tending towards a 2D magnetic behavior, since the intralayer exchange is the strongest. This is in contrast to the \GRS{} and \GRG{} systems and shows that electron doping due to Ag/Au substitution can also change the magnetic dimensionality. Moreover, the interlayer exchange along the [111] direction is now antiferromagnetic instead of ferromagnetic, while the on-site magnetic anisotropy is easy-plane for GdAg$_2$Si$_2$ and GdAu$_2$Si$_2$ and uniaxial for GdAg$_2$Ge$_2$ and GdAu$_2$Ge$_2$. As a direct consequence of all these differences, the ground-state spin configurations of these Au/Ag-based systems are distinctly different from what we discussed in the previous section for the \GRS{} and other 122-type compounds.

For example, GdAg$_2$Si$_2$, GdAu$_2$Si$_2$ and GdAu$_2$Ge$_2$ systems reveal a clear tendency towards antiferromagnetic stripe (Fig.~\ref{f:Au_Ag_systems}a) and non-collinear order, where the N\'eel vectors in the neighboring unit cells are 90\textdegree-rotated with respect to each other (Fig.~\ref{f:Au_Ag_systems}b). In contrast, GdAg$_2$Ge$_2$ reveals stability of individual topological textures with a diameter of just a few nanometers (Fig.~\ref{f:Au_Ag_systems}c) over a wide range of applied magnetic field. These textures are partially reminiscent of antiskyrmions. If we look at the parameters of the spin Hamiltonian in Table~I for these 4 compounds, it seems that the on-site anisotropy is not decisive for this unique behavior of GdAg$_2$Ge$_2$, which is more likely to be linked to the dramatically different ratio between nearest- and next-nearest-neighbor Heisenberg exchange interactions within each Gd layer ($J_{100}$ and $J_{110}$ in \-Table~I). For~GdAg$_2$Ge$_2$, $J_{100}$ is almost negligible compared to $J_{110}$, while the other 3 systems show comparable magnitudes for both exchange interactions. One should keep in mind that there are, of course, more exchange interaction paths due to the long-range character of Heisenberg interactions related to possible RKKY mechanism in this kind of systems.

Also, it should be mentioned that dipolar interactions are included in all these magnetic simulations and we find that they contribute significantly to the predicted magnetic ground state. For example, the AFM skyrmions in Fig.~\ref{f:Au_Ag_systems}c are not stabilized in absence of dipolar interactions, and the shape of the other 2 spin configurations (Fig.~\ref{f:Au_Ag_systems}a,b) can differ somewhat as well. The simple reason is that the large Gd moments enhance the dipolar interactions, allowing it to compete with relatively weak Heisenberg exchange.

\section{1144-type compounds}

Having discussed some trends in the 122-type compounds, we proceed now with the new 1144-type systems that we propose here. The idea is to replace every second Gd layer with non-magnetic alkali metal layer (\textit{A} = K, Rb or Cs), as shown in Fig.~\ref{f:1144_textures}d. Based on literature reports on iron-based superconductors with similar structure [\onlinecite{Iyo2016,Borisov2017,Borisov2018,Song2021}], it should be possible to stabilize this 1144-type structure experimentally, also for the considered Gd-based magnets. The important factor is the large difference between the ionic radii of Gd and the alkali elements K, Rb and Cs, which favors the perfectly ordered/separated Gd and alkali layers, instead of random distribution of these elements in different layers. For comparison, however, we also consider GdYRu$_4$Si$_4$ where Gd and Y are cations of similar size, and the valence state of Y is very different from K, Rb and Cs, so electron count effects will be important. Although alternating Gd and Y layers are not expected in this particular compound, for simplicity of comparison with the rest of the systems we will still assume a simple ordered pattern of Gd and Y layers, just to focus on the effects of electron count.

\begin{figure}
{\centering
\includegraphics[width=0.45\textwidth]{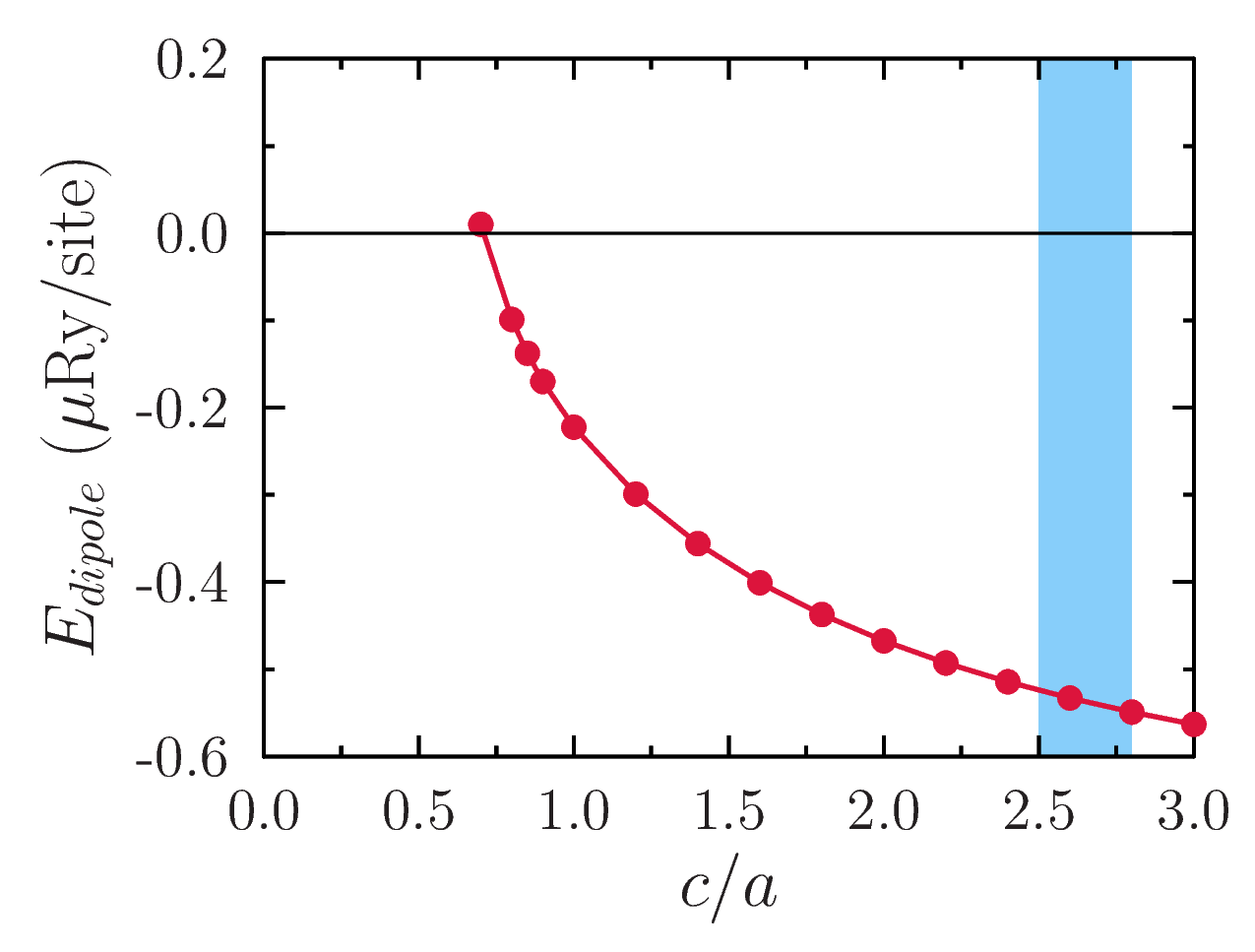}
}
\vspace{-10pt}
\caption{Average dipolar energy, calculated using Eqn.~\ref{e:dipolar_exchange}, per magnetic site in the 122-type crystal structure (Fig.~\ref{f:GdRu2Ge2}, middle plot) as a function of the tetragonality ratio $c/a$. The range of $c/a$ values obtained for the 1144-type systems considered in this work is indicated by the light-blue-shaded region.}
\vspace{-5pt}
\label{f:Edip_vs_ctoa}
\end{figure}

\begin{figure*}
\vspace{10pt}
{\centering
\includegraphics[width=0.95\textwidth]{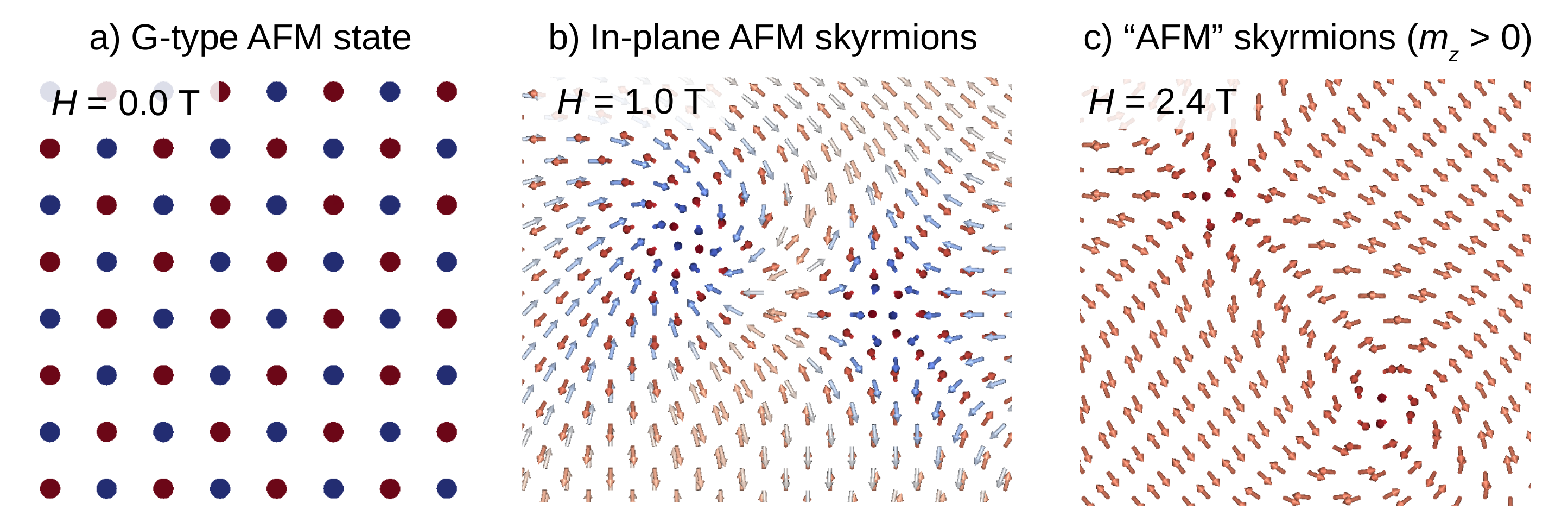}
\includegraphics[width=0.95\textwidth]{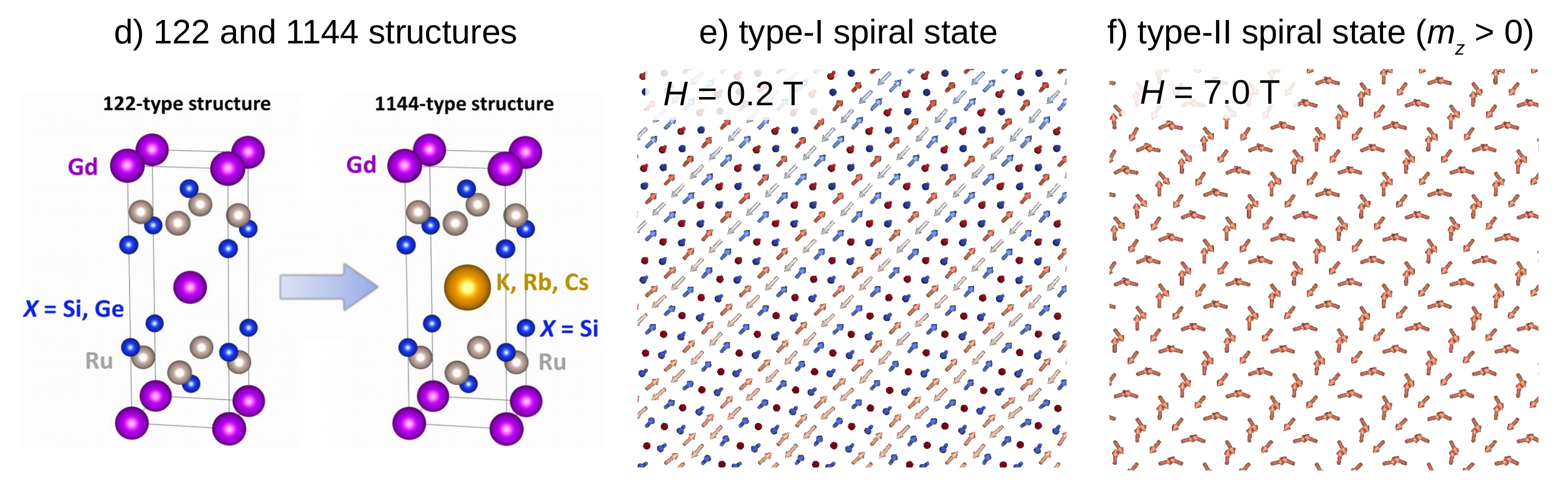}
\includegraphics[width=0.95\textwidth]{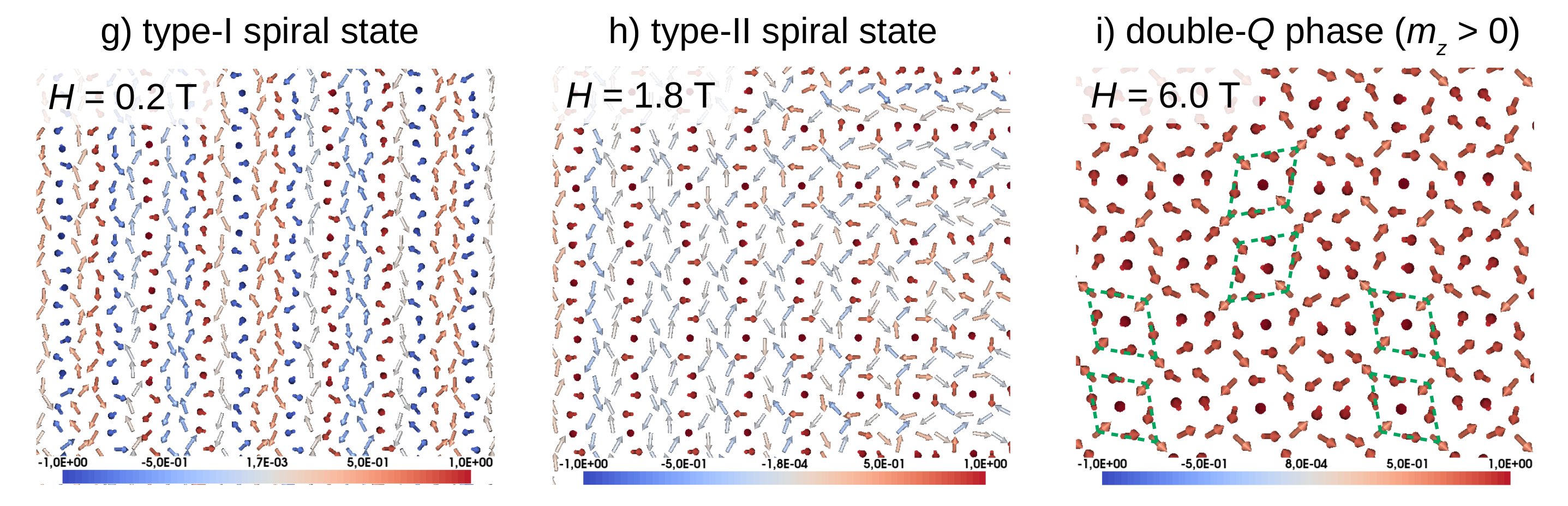}
}
\caption{Simulated field-dependent zero-temperature spin configurations for the 1144-type compounds a-c) GdKRu$_4$Si$_4$, e,f) GdCsRu$_4$Si$_4$ (GdRbRu$_4$Si$_4$ is similar), and g-i) GdYRu$_4$Si$_4$; the color code corresponds to the $z$-component of atomic spin. The difference between the 122- and 1144-type crystal structures is illustrated by plot d). In the 1144 compounds, the Gd and alkali metal layers are alternating and perfectly ordered. In case of GdYRu$_4$Si$_4$, the calculations assume the same 1144-type structure, despite similar ionic sizes of Gd and Y.}
\label{f:1144_textures}
\end{figure*}

\textbf{Doping effects.} There are 3 obvious consequences of alkali metal substitution in GdRu$_2$Si$_2$, which can be deduced without any calculation. Firstly, the interlayer magnetic interaction $J_{111}$, which is significant in GdRu$_2$Si$_2$, (see Tables~I and II) is not present anymore, because every second Gd is replaced by non-magnetic alkali metal. Already based on this, one can expect weaker magnetism in derived Gd\textit{A}Ru$_4$Si$_4$ compounds. Secondly, depending on the ionic size of the alkali cation, the spacing between neighboring Gd layers can change significantly, which directly influences the dipolar interactions in these quasi-2D systems according to Eqn.~(\ref{e:dipolar_exchange}). However, our model calculations using supercells of varying size have shown that the dipole energy of the whole system is not very sensitive to the $c/a$ ratio for the range of its values that we get for the studied 122- and 1144-type systems (Fig.~\ref{f:Edip_vs_ctoa}). It would be sensitive to $c/a$ for systems close to the cubic structure where $c/a$ is nominally equal to $1/\sqrt{2}$. Finally, introduction of alkali cations will reduce the electron count of Ru sites by 0.25~electrons per Ru, reshaping thereby the Fermi surface (FS) and RKKY interactions. In fact, the calculated FS's of GdRu$_2$Si$_2$ and GdRu$_2$Ge$_2$ are quite similar, while the Fermi surface of e.g.~GdKRu$_4$Si$_4$ (Fig.~\ref{f:Fermi_surfaces}e), is distinctly different, showing less FS sheets and weaker dispersion of electronic bands in the $z$-direction (the later is due to increased $c/a$ ratio). Interestingly, the inner hole pockets are not much affected by the change of electron count for GdKRu$_4$Si$_4$; most changes are observed for the outer FS sheets closer to the Brillouin zone boundaries.

\textbf{Effective spin model.} Let us discuss the magnetic trends in the new 1144-type Gd\textit{A}Ru$_4$Si$_4$ compounds (\textit{A} = K, Rb, Cs). The calculated Heisenberg interactions are clearly different from GdRu$_2$Si$_2$ (see Table~I), showing strongly ferromagnetic diagonal exchange $J_{110}$, while $J_{100}$ exchange is still AFM. At first glance, this should lift the magnetic frustration. However, there are also significant further-neighbor interactions which maintain some degree of frustration in the system. Overall, the energy scale of Heisenberg interactions decreases in the K-Rb-Cs series, meaning that also the interlayer exchange $J_{001}$ is weaker and the 1144-type systems are more two-dimensional than the 122-type systems. The weakening of $J_{001}$ is especially significant due to increased interlayer spacing induced by larger Cs cations (see data on $d^{(2)}_{X-X}$ in Table~I). Also, now the $J_{111}$ exchange is zero, since alkali metal cations are non-magnetic, leaving $J_{001}$ to be the nearest-neighbor exchange between Gd layers. Another important modification compared to GdRu$_2$Si$_2$ is the easy-plane on-site anisotropy ($K_U > 0$) for GdK- and GdRb-based compounds, instead of uniaxial anisotropy in GdRu$_2$Si$_2$ ($K_U < 0$).

\textbf{Numerical issues.} When it comes to simulating atomistic spin dynamics, there is a following numerical issue with these 1144-type systems due to the weak interlayer exchange (e.g.~$J_{001}$). If the simulation starts with a random spin configuration, the equilibrated magnetic state at zero temperature might show different spin alignments between the neighboring layers, which differ by only a small energy due to a weak $J_{001}$. Many such local energy minima can be anticipated due to the complexity of Heisenberg interactions, making it difficult to find the actual ground state. In this work, we restrict the search for the equilibrium magnetic configuration by imposing the same spin configuration in each Gd layer, which is done by choosing the $z$-dimension of the simulation supercell to be equal to one chemical unit cell. This assumption is based on the fact that the calculated interlayer exchange $J_{001}$ favors ferromagnetic alignment for all 1144-type systems considered here. While this simplifying assumption may drift the calculations away from the actual ground state, it makes the calculations numerically more stable and allows to get a more clear picture of magnetic ordering tendencies within each Gd layer, where magnetic interactions are the strongest anyway.

\textbf{Magnetic textures.} As a result of field-dependent ASD simulations with the calculated microscopic magnetic parameters and including the dipolar interactions, we identify a variety of magnetic phases in the proposed Gd\textit{A}Ru$_4$Si$_4$ compounds (\textit{A} = K, Rb, Cs), which are distinct from the phases observed for the GdRu$_2$Si$_2$ and other systems discussed in the previous sections. For example, for GdRbRu$_4$Si$_4$ and GdCsRu$_4$Si$_4$ we find a 45\textdegree-rotated spin-spiral state (Fig.~\ref{f:1144_textures}e), corresponding to a $\vec{Q}_{110}$ ordering vector, which transforms into another spiral state with finite polarization along the field direction at around 1--2\,T (Fig.~\ref{f:1144_textures}f) and then turns into a fully polarized state at fields around 9--12\,T. In contrast, the GdKRu$_4$Si$_4$ compound is collinear antiferromagnet with moments oriented out-of-plane and forming a checkerboard arrangement within each Gd layer (Fig.~\ref{f:1144_textures}a). Above an applied magnetic field of 1\,T, it transforms into primarily an in-plane-oriented antiferromagnet with individual skyrmion-like objects (Fig.~\ref{f:1144_textures}b,c), which are easily identifiable especially from the real-space distribution of the topological charge (Figs.~\ref{f:topological_charge_1144}a-d). These localized skyrmion-like spin textures do not show a tendency towards forming ordered lattices in GdKRu$_4$Si$_4$, as opposed to skyrmions in GdRu$_2$Si$_2$ and GdRu$_2$Ge$_2$ compounds. However, they often appear in pairs in our spin dynamics simulations, depending on the initial random spin configuration, and bear similarity to meron-antimeron pairs discussed in the literature for other systems [\onlinecite{Yu2018},\onlinecite{Gao2019}]. To understand the fundamental difference between the K and Rb/Cs systems, we performed additional calculations to find out the role of different terms in the Hamiltonian (\ref{e:spin_model}) as well as the dipolar exchange (\ref{e:dipolar_exchange}). Interestingly, setting the on-site anisotropy or dipolar exchange to zero does not destabilize the aforementioned localized spin textures in GdKRu$_4$Si$_4$, suggesting that the Heisenberg exchange must be the source of their stability.

\begin{figure}
{\centering
\includegraphics[width=0.45\textwidth]{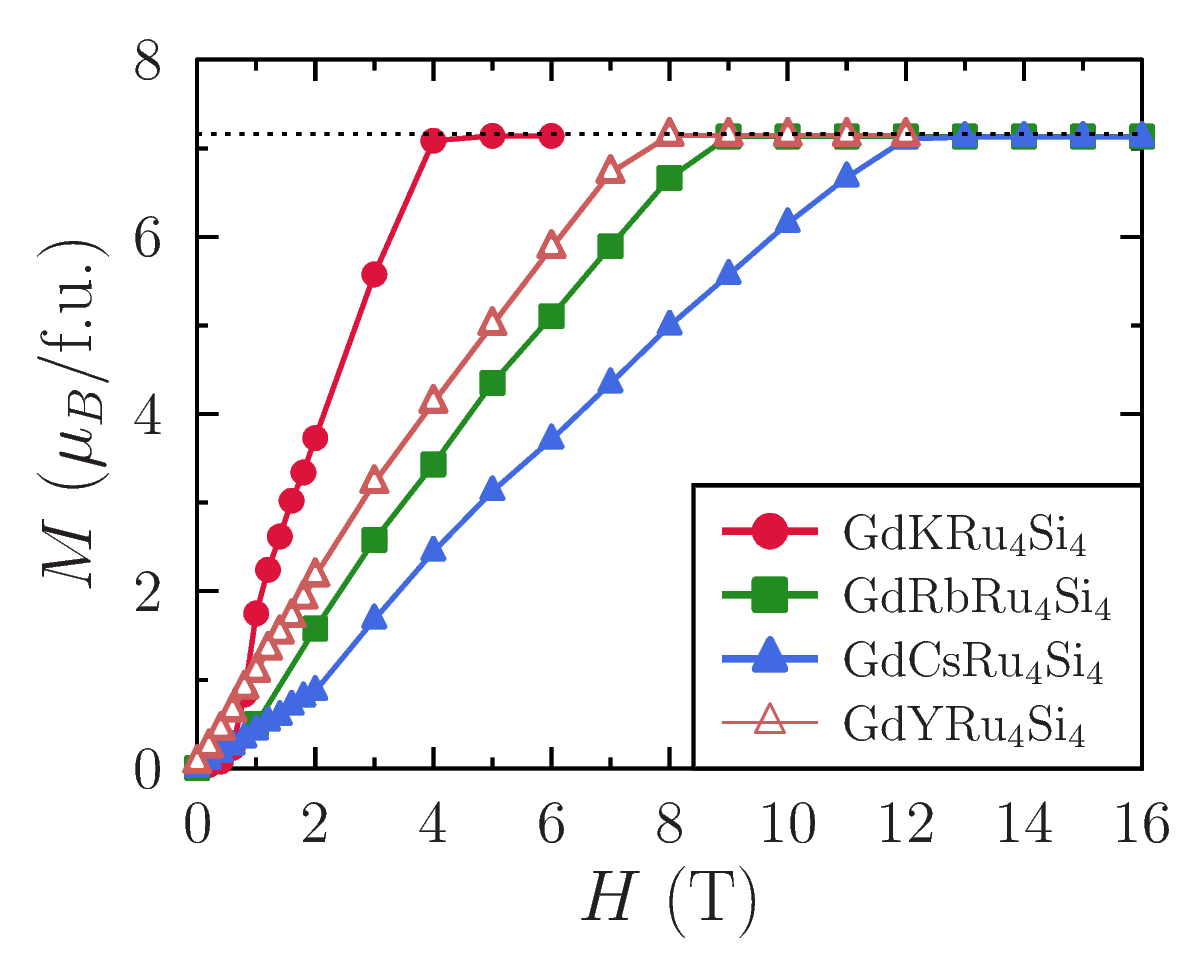}
}
\vspace{-10pt}
\caption{Simulated magnetization vs applied field for the 1144-type Gd\textit{A}Ru$_4$Si$_4$ compounds (\textit{A} = K, Rb, Cs, Y).}
\label{f:M_vs_H_1144}
\end{figure}

If we also look at the magnetization vs applied field curves shown in Fig.~\ref{f:M_vs_H_1144}, we can notice the increasing saturation field, at which the system becomes fully spin-polarized, in the series K-Rb-Cs. This cannot be attributed to the on-site anisotropy; if we set the on-site anisotropy for GdCsRu$_4$Si$_4$ to the value obtained for GdKRu$_4$Si$_4$, then the saturation field for GdCsRu$_4$Si$_4$ decreases only slightly, which is not enough to explain the dramatic difference between the two compounds, allowing us to conclude that the magnetization curves are mostly determined here by the Heisenberg exchange interactions.

Alongside the 3 hypothetical compounds discussed above, we have also considered GdYRu$_4$Si$_4$ which is expected to have a very similar electron count to a pure Gd-based compound, due to Y and Gd both being trivalent. This is also immediately apparent from the Fermi surfaces (Fig.~\ref{f:Fermi_surfaces}), where GdYRu$_4$Si$_4$ looks similar to GdRu$_2$Si$_2$ due to closer electron count. Also the Heisenberg magnetic interactions and on-site anisotropy are quite different from the K-, Rb- and Cs-based systems. From the atomistic spin dynamics simulations it follows that the low-field ground states are spin spirals of different types, while complex double-$\vec{Q}$ magnetic state is stabilized at higher applied field (Fig.~\ref{f:1144_textures}g-i); the double-$\vec{Q}$ character is more clearly illustrated by the topological charge in Fig.~\ref{f:topological_charge_1144}n. These data, together with results in Section~IV, indicate the vast opportunities for tuning the magnetic  behavior of layered rare-earth magnetic systems with ThCr$_2$Si$_2$-type structure through substitution of different interlayer cations (Y, K, Rb, Cs) and transition metals (Ru, Ag, Au).

It should be noted that the discussed results for the 1144-type systems are obtained for theoretically optimized crystal structures, since Gd\textit{A}Ru$_4$Si$_4$ compounds have not been synthesized yet, so no measured structures are available so far. Since magnetic interactions can be sensitive to structural details, as we demonstrated on the example of GdRu$_2$Si$_2$, our theory predictions for Gd\textit{A}Ru$_4$Si$_4$ might also change when recalculated for the experimental structures, once they are available. In this respect, we note that our recent calculations for GdRu$_2$Si$_2$ [\onlinecite{Sarkar2026}] showed that the optimized crystal structure and magnetic parameters calculated within the GGA-PBE approximation reproduce the measured magnetic phase diagram satisfactorily. This justifies the application of this approximation in the present work and speaks in favor of reliability of theory predictions for the different 122- and 1144-type systems considered in this work. The presented comparative analysis of 122- and 1144-systems strongly suggests that the magnetic properties of 1144-systems can be qualitatively different from the 122-systems, motivating future experimental studies.

\section{Summary}

Theoretical modelling on different length scales using electronic structure theory and atomistic spin dynamics have revealed here several curious trends and new insights for layered intermetallic compounds based on GdRu$_2$Si$_2$.

The first observation is that Ge substitution of Si changes magnetic interactions visibly but, despite this, leaves the magnetic phase diagram qualitatively the same, which partially agrees with experiment. However, we find that Heisenberg interactions in this kind of magnets are quite sensitive, e.g., to the Si position and lattice parameters, leading to challenges in theoretical modelling. Also, dipolar interactions appear to be crucial for the skyrmion stability in GdRu$_2$Si$_2$ and similar systems on the nanometer length scale, which is quite unique. One major issue, which is yet to be understood, is the fact that simulations predict higher saturation field for GdRu$_2$Ge$_2$ than for GdRu$_2$Si$_2$, while experiment actually shows the opposite. This can spur further interest in these two skyrmionic systems.

Secondly, for systems where Ru is replaced by Au or Ag we find a tendency towards a topologically trivial antiferromagnetic order (Fig.~\ref{f:Au_Ag_systems}a,b), although skyrmion-like textures on the nanometer length scale can be stabilized at certain applied field. The latter is especially true for GdAg$_2$Ge$_2$. This might be related to the large ratio between the nearest- and next-nearest-neighbor Heisenberg interactions, which sets this system apart from the other 3 Ag/Au-based compounds.

Finally, we propose the possibility of chemical tuning of GdRu$_2$Si$_2$ by alkali metals which, due to their much larger ionic radii, are very likely to form separate layers alternating with Gd layers, similarly to well-known iron pnictide superconductors like CaKFe$_4$As$_4$. We predict the resulting Gd\textit{A}Ru$_4$Si$_4$ (\textit{A} = K, Rb, Cs) compounds to host a variety of spin-spiral phases which, for GdRb- and GdCs-systems, are similar to the spin configurations observed in GdRu$_2$Si$_2$. In complete contrast to this, the GdK-system starts with a checkerboard AFM order at zero field and transforms into in-plane-polarized domains with individual skyrmion-like localized spin textures. Such dramatic changes in the magnetic properties are not surprising in view of significantly modified magnetic interactions in Gd\textit{A}Ru$_4$Si$_4$ compared to GdRu$_2$Si$_2$. Interestingly, dipolar energy is less important in Gd\textit{A}Ru$_4$Si$_4$ for the topological magnetism, even though the system is more two-dimensional due to Gd layers separated by non-magnetic alkali metal layers, while the role of frustrated Heisenberg interactions is increased.

Furthermore, substitution with Y instead of alkali-earth metals provides further opportunities for dramatic tuning of magnetic properties (Fig.~\ref{f:1144_textures}g-i), since trivalent Y changes the Fermi surface and consequently magnetic interactions mediated by conduction electrons, compared to Gd\textit{A}Ru$_4$Si$_4$ with monovalent \textit{A} species. Interestingly, microscopic magnetic parameters for GdYRu$_4$Si$_4$ and GdRu$_2$GeSi (see~Table~I) are mostly similar, which can be attributed to similar electron counts and Fermi surfaces of these 2 compounds, further supporting the idea that the magnetic properties are governed by conduction electrons and RKKY-like mechanism.

Overall, based on these results, we expect rich opportunities for chemical tuning of Gd-based layered magnets, starting with the prototype GdRu$_2$Si$_2$ compound, with a goal to design new interesting systems with non-collinear or even topologically non-trivial magnetic phases on the nanometer length scale not requiring chiral interactions, which presents a purely fundamental interest.

\section{Data availability}

All magnetic interactions and other parameters of Hamiltonian (\ref{e:spin_model}) calculated in this work as well as crystal structures used in these calculations are available from the corresponding author upon reasonable request. Theoretical data have been obtained using publicly available full-potential electronic structure RSPt [\onlinecite{Wills1987},\onlinecite{Wills2010}] and atomistic spin dynamics UppASD [\onlinecite{uppasd},\onlinecite{Eriksson2017}] softwares and commercially available VASP code [\onlinecite{kresse1996efficient}].

\section{Acknowledgements}

This work was financially supported by the Swedish Research Council through grant number 2024.05206 (PI: V.B.) and Knut and Alice Wallenberg Foundation through grant numbers 2018.0060, 2021.0246, and 2022.0108 (PI's: O.E.~and A.D.). R.P.~and V.B.~were supported by the G\"oran Gustafsson Foundation (recipient of the ``small prize'': V.B.). O.E.~and A.D.~acknowledge support from the Wallenberg Initiative Materials Science for Sustainability (WISE) funded by the Knut and Alice Wallenberg Foundation (KAW). A.D. also acknowledges financial support from the Swedish Research Council (Vetenskapsrådet, VR), Grant No. 2016-05980 and Grant No. 2019-05304. O.E.~also acknowledges support by the Swedish Research Council (VR), the Foundation for Strategic Research (SSF), the Swedish Energy Agency (Energimyndigheten), the European Research Council (854843-FASTCORR), eSSENCE and STandUP. S.S.~acknow\-ledges funding (postdoctoral stipend) from the Carl Tryggers Foundation (grant number CTS 22:2013, PI: V.B.).

The computations/data handling were enabled by resources provided by the National Academic Infrastructure for Supercomputing in Sweden (NAISS) at the National Supercomputing Centre (NSC, Tetralith cluster) partially funded by the Swedish Research Council through grant agreement no.\,2022-06725. Structural sketches in Figs.~\ref{f:GdRu2Ge2}, \ref{f:GdRu2GeSi} and \ref{f:1144_textures} were produced using the \textsc{VESTA3} software \cite{vesta}. Spin configurations in Figs.~\ref{f:GdRu2Ge2}, \ref{f:GdRu2GeSi}, \ref{f:Au_Ag_systems}, \ref{f:1144_textures}, \ref{f:topological_charge_GdRu2Si2}, \ref{f:topological_charge_GdRu2Ge2} and \ref{f:topological_charge_1144} were plotted using the UppASD graphical user interface \cite{uppasd}. The Fermi surfaces in Fig.~\ref{f:Fermi_surfaces} were plotted using the XCrysden software [\onlinecite{xcrysden}].

\medskip
\bibliographystyle{prb-titles.bst}
\bibliography{main}

\appendix
\section{Magnetic interactions and on-site anisotropy for all studied compounds}

Table~I further below summarizes some of the nearest-neighbor Heisenberg interactions and on-site anisotropy in 122- and 1144-type Gd-based systems. It should be noted that we have calculated the magnetic interactions for hundreds of neighbors, and they show a long-range character, meaning that not only the first neighbors are important. For discussing some basic trends across the studied compounds, however, it is still useful to look at the nearest-neighbor data shown in this table.

\setlength{\tabcolsep}{5pt}
\renewcommand{\arraystretch}{1.5}
\setlength{\tabcolsep}{7pt}
\begin{table*}
 \caption{Heisenberg exchange parameters for a few nearest neighbors shown in Fig.~\ref{f:GdRu2Ge2} ($J<0$ means antiferromagnetic interaction, $J>0$ is ferromagnetic) and uniaxial anisotropy constant ($K_U<0$ means that the $z$-direction is the easy axis, $K_U>0$ corresponds to $xy$ easy-plane anisotropy) calculated within density functional theory (DFT) for different 122- and 1144-type systems. The lattice parameters $a$ and $c$ as well as the interlayer distance $d_{X-X}$ (in {\AA}) between the neighboring Si or Ge ions are shown as well. Optimized structures are obtained by minimizing Hellmann-Feynman forces calculated using DFT. The $J$ and $K_U$ values are given in units of $\mu\mathrm{Ry}$ (1\,$\mu\mathrm{Ry} \approx \unit[0.16]{K}$).}
\vspace{5pt}
  \centering
  \begin{tabular}{c|cc|c|c}
    \hline
            & \multicolumn{2}{c|}{experimental structures}   & interpolated &  optimized structure \\
              & GdRu$_2$Si$_2$ & GdRu$_2$Ge$_2$ & GdRu$_2$GeSi & GdRu$_2$GeSi \\
    \hline
    $J_{100}$ & $-3.7$ & $-8.9$ & $-5.6$ & $+5.8$  \\
    $J_{110}$ & $-12.4$ & $-27.5$ & $-18.6$ & $-13.7$  \\
    $J_{111}$ & $+94.6$ & $+62.9$ & $+74.0$; $+65.0$ & $+73.6$; $+65.4$  \\
    $J_{001}$ & $+21.3$ & $+20.7$ & $+20.1$ & $+15.9$ \\
    $K_U$ & $-3.8$ & $-0.41$ & $-1.6$ & $-2.1$  \\
    \hline
    $a$ (\AA)       & $4.16$ & $4.24$ & $4.19$ & $4.21$ \\
    $c$ (\AA)       & $9.61$ & $9.90$ & $9.81$ & $9.85$ \\
    $d_{X-X}$ (\AA) & $2.40$ & $2.24$ & $2.34$ & $2.58$ \\
    \hline
  \end{tabular}

  \vspace{10pt}
  \begin{tabular}{c|ccccc}
    \hline
              & \multicolumn{4}{c}{optimized structures} \\
              & GdAg$_2$Si$_2$ & GdAg$_2$Ge$_2$ & GdAu$_2$Si$_2$ & GdAu$_2$Ge$_2$ \\
    \hline
    $J_{100}$       & $-24.3$ &   $-3.6$ & $-51.5$ & $-47.3$ \\
    $J_{110}$       & $-89.0$ & $-110.4$ & $-56.7$ & $-65.8$ \\
    $J_{111}$       &  $-7.4$ &  $-10.1$ & $-19.7$ & $-15.7$ \\
    $J_{001}$       &  $+4.6$ &   $+8.3$ & $+5.3$ & $+4.2$ \\
    $K_U$           & $+1.1$ & $-0.21$ & $+0.26$ & $-0.72$ \\
    \hline
    $a$ (\AA)       &  $4.18$ &  $4.24$ &  $4.28$ &  $4.40$ \\
    $c$ (\AA)       & $10.78$ & $11.33$ & $10.27$ & $10.58$ \\
    $d_{X-X}$ (\AA) &  $2.30$ &  $2.44$ &  $2.31$ &  $2.44$ \\
    \hline
  \end{tabular}
\end{table*}
\begin{table*}

\vspace{5pt}
  \centering
  \begin{tabular}{c|cccc}
    \hline
               &  \multicolumn{4}{c}{optimized structures} \\
               & GdYRu$_4$Si$_4$ & GdKRu$_4$Si$_4$ & GdRbRu$_4$Si$_4$ & GdCsRu$_4$Si$_4$ \\
    \hline
    $J_{100}$  &  $+8.2$ & $-14.6$ & $-12.8$ & $-14.7$ \\
    $J_{110}$  & $-12.8$ & $+47.8$ & $+35.8$ & $+27.8$ \\
    $J_{111}$  & $-$ & $-$ & $-$ & $-$ \\
    $J_{001}$  & $+15.6$ & $+8.4$ & $+3.6$ & $+1.6$  \\
    $K_U$      &  $-2.3$ & $+2.6$ & $+2.6$ & $-0.86$ \\
    \hline
    $d^{(1)}_{X-X}$  & $2.56$ & $2.58$ & $2.58$ & $2.58$  \\
    $d^{(2)}_{X-X}$  & $2.53$ & $3.83$ & $4.49$ & $5.06$  \\
    \hline
  \end{tabular}
\end{table*}

\section{Phonon and magnon spectra of Ru-based compounds}

\begin{figure*}
{\centering
\includegraphics[width=0.99\textwidth]{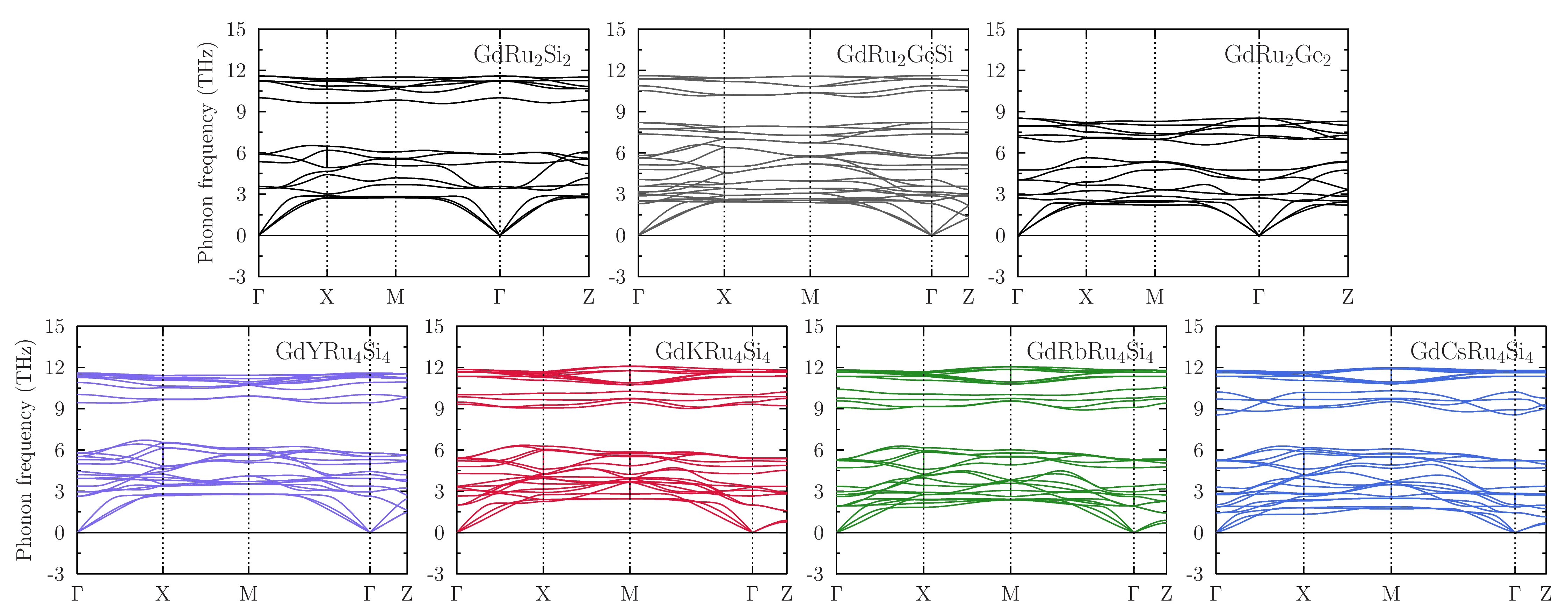}
}
\vspace{-10pt}
\caption{Adiabatic phonon spectra calculated for the 122-type GdRu$_2$Si$_{2-x}$Ge$_x$ ($x=0,1,2$) and 1144-type Gd\textit{A}Ru$_4$Si$_4$ compounds (\textit{A} = Y, K, Rb, Cs). All the studied compounds show no imaginary phonon frequencies, indicating dynamical stability.}
\vspace{-10pt}
\label{f:phonons}
\end{figure*}

To verify the dynamical stability of the studied 122- and 1144-type compounds, we have calculated the adiabatic phonon spectra (Fig.~\ref{f:phonons}) using density functional theory. These first-principles calculations were performed using the \textsc{vasp} (\textit{Vienna Ab-initio Simulation Package})~[\onlinecite{kresse1996efficient}] in combination with \textsc{phonopy}~[\onlinecite{togo2015phonon}] to obtain interatomic force constants and phonon band structures, focusing on the compounds $\mathrm{GdRu_{2}Si_{2}}$, $\mathrm{GdRu_{2}GeSi}$, $\mathrm{GdRu_{2}Ge_{2}}$, $\mathrm{GdYRu_{4}Si_{4}}$, $\mathrm{GdKRu_{4}Si_{4}}$, $\mathrm{GdRbRu_{4}Si_{4}}$, and $\mathrm{GdCsRu_{4}Si_{4}}$.

Spin-polarized density functional theory (DFT) calculations were carried out using the projector augmented-wave (PAW) method~[\onlinecite{kresse1999ultrasoft}], as implemented in \textsc{vasp}~[\onlinecite{kresse1996efficient}]. The exchange–correlation functional was treated within the generalized gradient approximation (GGA) using the Perdew–Burke–Ernzerhof (PBE) parametrization~[\onlinecite{perdew1996generalized}].

A plane-wave energy cutoff of $E_{\mathrm{cut}} = 600$~eV was employed, and the Brillouin zone was sampled using a $\Gamma$-centered Monkhorst–Pack $20 \times 20 \times 8$ $k$-mesh to ensure convergence of the total energy and local magnetic moments. To properly account for the localized nature of the Gd~4$f$ electrons, the rotationally invariant DFT+$U$ scheme introduced by Liechtenstein \textit{et al.}~[\onlinecite{liechtenstein1995density}] was used, with on-site Coulomb and exchange parameters of $U = 7.0$~eV and $J = 1.05$~eV, respectively.

\begin{figure*}
{\centering
\includegraphics[width=0.99\textwidth]{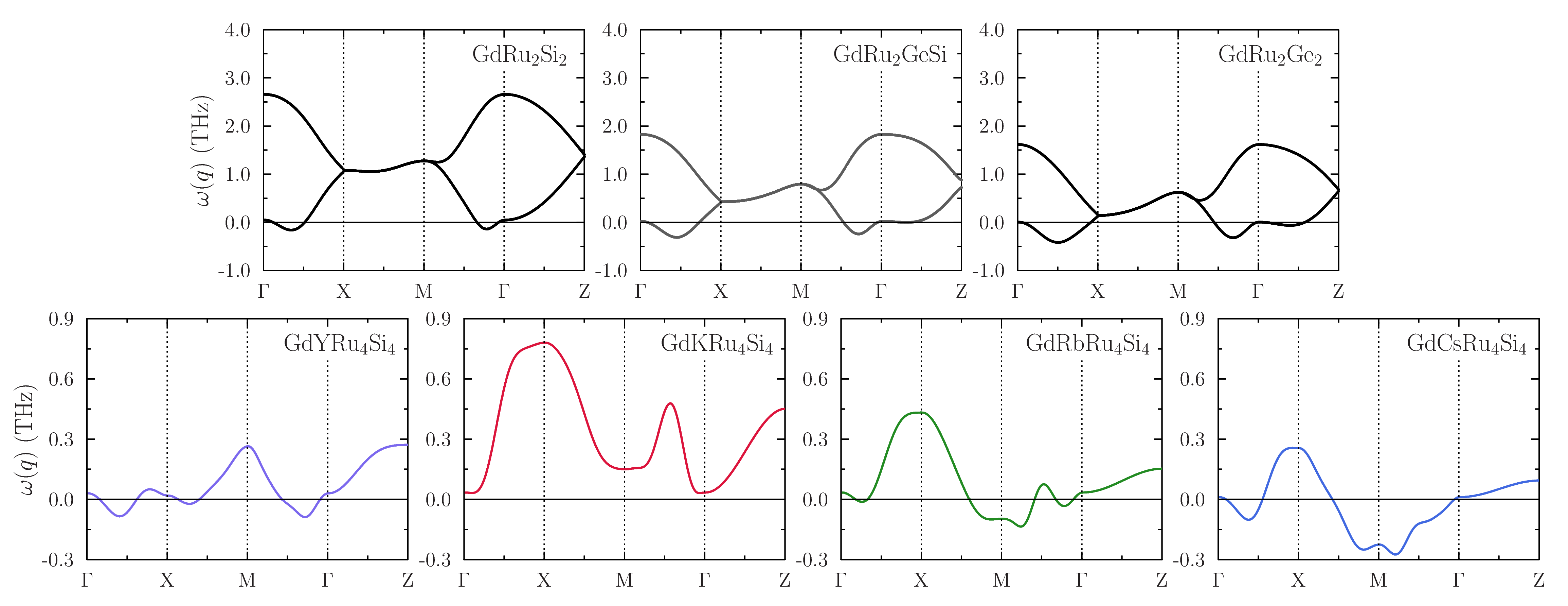}
}
\vspace{-10pt}
\caption{Adiabatic magnon spectra calculated for the 122-type GdRu$_2$Si$_{2-x}$Ge$_x$ ($x=0,1,2$) and 1144-type Gd\textit{A}Ru$_4$Si$_4$ compounds (\textit{A} = Y, K, Rb, Cs). Negative values here actually correspond to imaginary magnon frequencies ($\omega^2 < 0$) indicating potential magnetic instability with respect to the reference ferromagnetic state.}
\vspace{-10pt}
\label{f:magnons}
\end{figure*}
For each compound, the 10-atom tetragonal conventional cell was fully relaxed with respect to atomic positions, cell shape, and volume until the Hellmann–Feynman forces on all atoms were less than 1~meV/Å. The interatomic force constants were then obtained using density functional perturbation theory (\textsc{dfpt}), as available in \textsc{vasp}, and subsequently analyzed using \textsc{phonopy}~[\onlinecite{togo2015phonon}]. The $2 \times 2 \times 2$ supercell was constructed to capture some of the long-range interatomic interactions essential for accurate phonon dispersion relations. For these supercell calculations, a $4 \times 4 \times 1$ $k$-mesh was used to ensure convergence.

For most compounds, the phonon spectra are qualitatively similar and indicate dynamic structural stability, based on the absence of imaginary phonon frequencies. Interestingly, there is a clear energy gap between 2 manifolds of phonon modes around $\unit[6-9]{THz}$ for most of the studied compounds, except for GdRu$_2$Ge$_2$ where this gap is relatively small. The origin of this peculiar feature could be worth exploring in a future work.

Concerning the adiabatic magnon spectra there is a clear qualitative similarity among the 122 compounds (top plots in Fig.~\ref{f:magnons}) in terms of the $\omega(q)$ dispersion, its magnitude and presence of imaginary frequencies along the $\Gamma-X$ and $\Gamma-M$ paths in $\vec{q}$-space. This magnetic instability is the most pronounced for GdRu$_2$Ge$_2$ and is observed at a larger $q$-vector compared to GdRu$_2$Si$_2$ and GdRu$_2$GeSi. For the Gd\textit{A}Ru$_4$Si$_4$ compounds with \textit{A} = Y, K, Rb, Cs (bottom plots in Fig.~\ref{f:magnons}) we see that the on-site easy-plane anisotropy pushes the magnon frequencies towards higher real-number values, which removes magnetic instabilities for GdKRu$_4$Si$_4$ but not completely for GdRbRu$_4$Si$_4$. Interestingly, the remaining instability in the latter system is observed here around $M$ point, while the instability between $\Gamma-X$ is lifted due to the on-site anisotropy. The GdCsRu$_4$Si$_4$ compound with uniaxial anisotropy (Table~I) reveals strong magnon instabilities in the whole $q_x-q_y$-plane. Overall, the GdRb- and GdCs-based systems have similar magnon dispersions, but shifted relative to each other. This similarity is reflected in qualitatively identical magnetic ground state in applied field illustrated for GdRbRu$_4$Si$_4$ in Fig.~\ref{f:1144_textures}e,f.

The GdYRu$_4$Si$_4$ compound is quite different from the 3 other 1144-type systems, as shown in the bottom left plot in Fig.~\ref{f:magnons}. This is to be expected due to the different electron count, since the valence state of Y cation is 3+, compared to 1+ for K, Rb and Cs. As discussed in the main text, this changes significantly the topology of the Fermi surface (Fig.~\ref{f:Fermi_surfaces}) and Heisenberg magnetic interactions (Table~I) as well as switches the on-site anisotropy to the strongly uniaxial character. As a result, magnetic instability shifts away from the M point towards the regions between $\Gamma$-X and $\Gamma$-M. Remarkably, the magnitudes of Heisenberg exchange within Gd layers and between them in the [001] direction are comparable in this Y-based compound, while the other 1144 systems are clearly more two-dimensional in terms of their magnetic interactions. This is again similar to the behavior of 122 systems of GdRu$_2$Si$_2$, GdRu$_2$Ge$_2$ and GdRu$_2$GeSi, where the interlayer exchange is actually the strongest.

\section{Topological charge}

\begin{figure*}
{\centering
\includegraphics[width=0.47\textwidth]{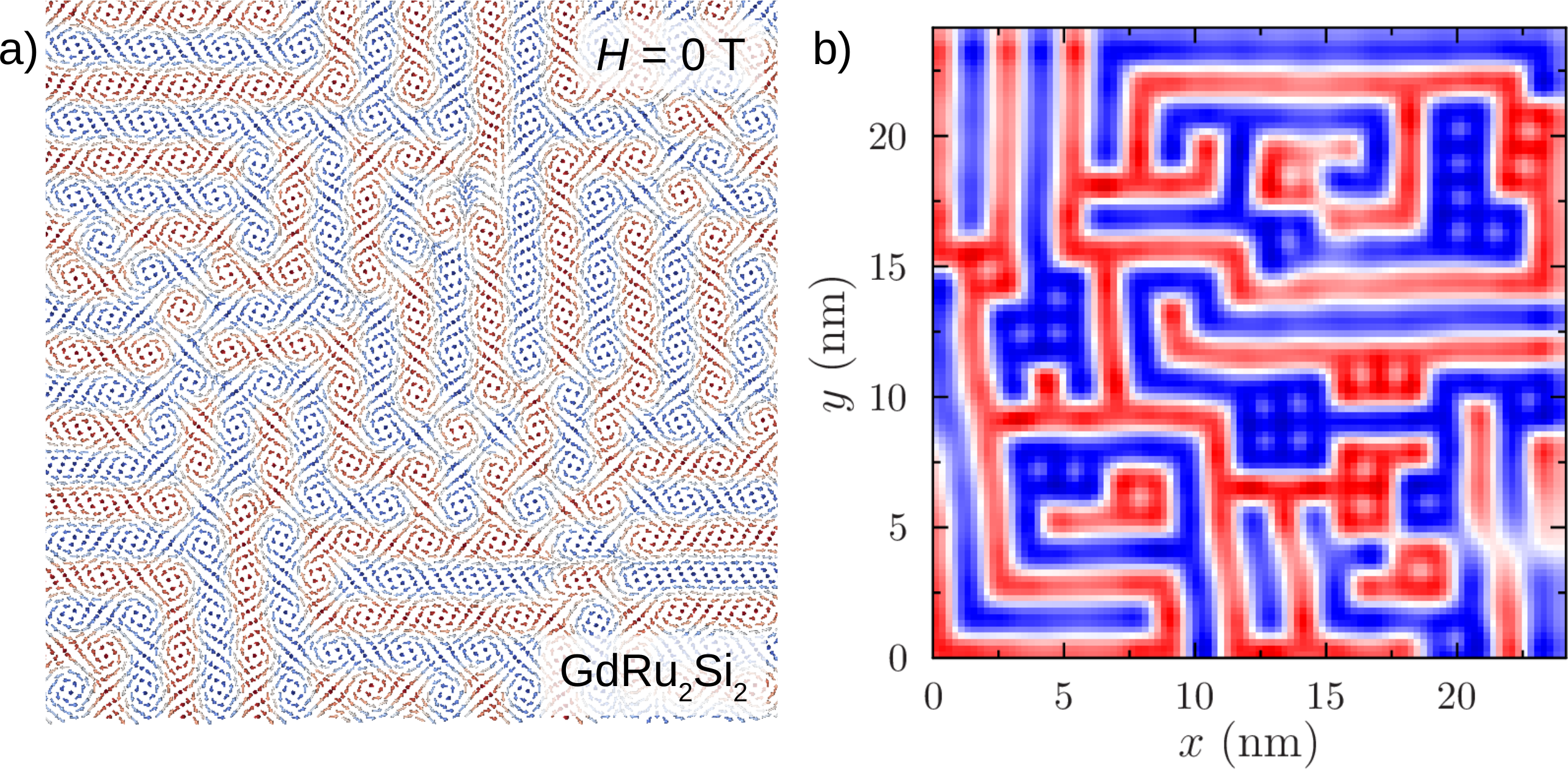}\hspace{10pt}
\includegraphics[width=0.47\textwidth]{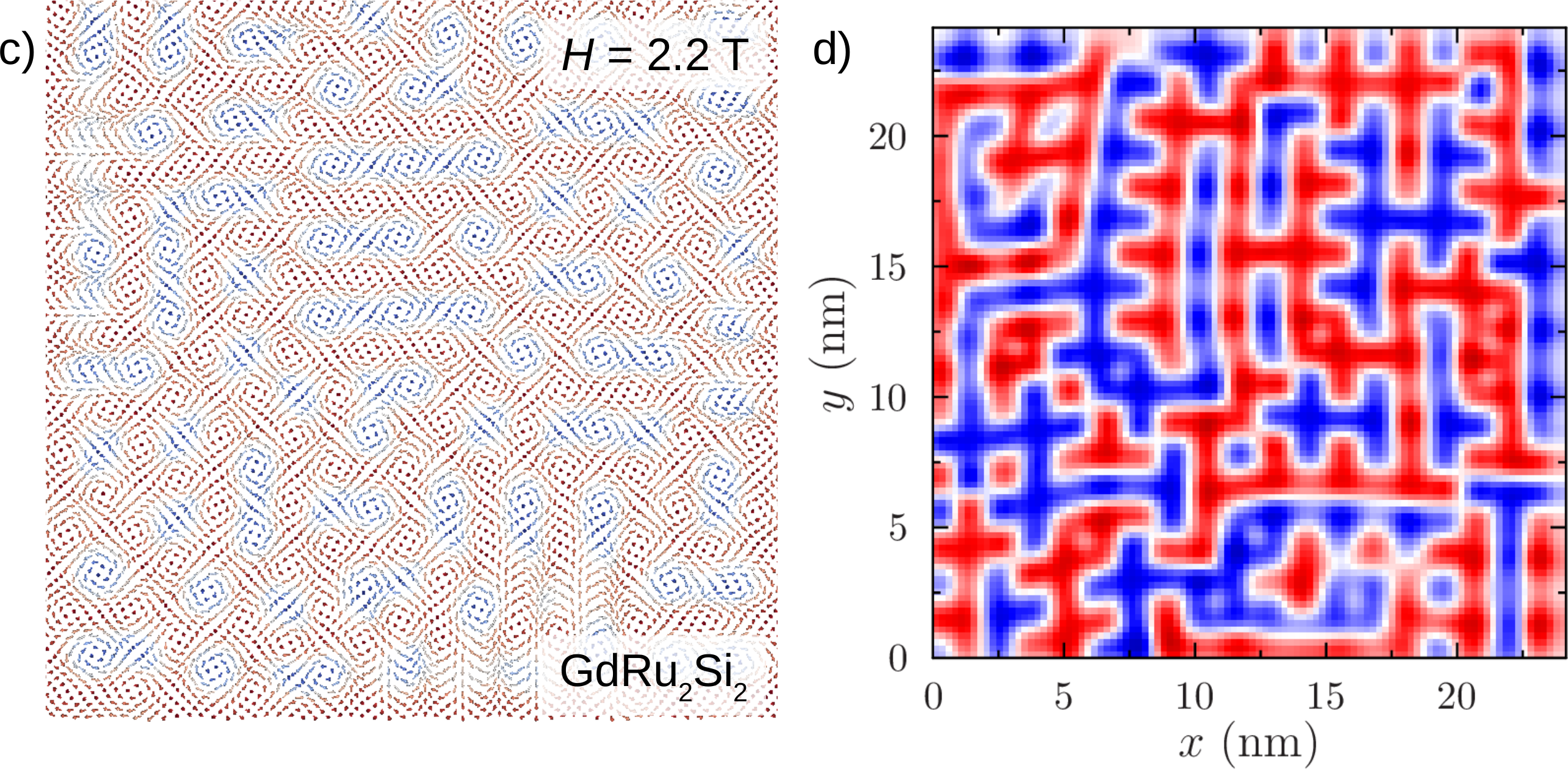}\\[10pt]
\includegraphics[width=0.47\textwidth]{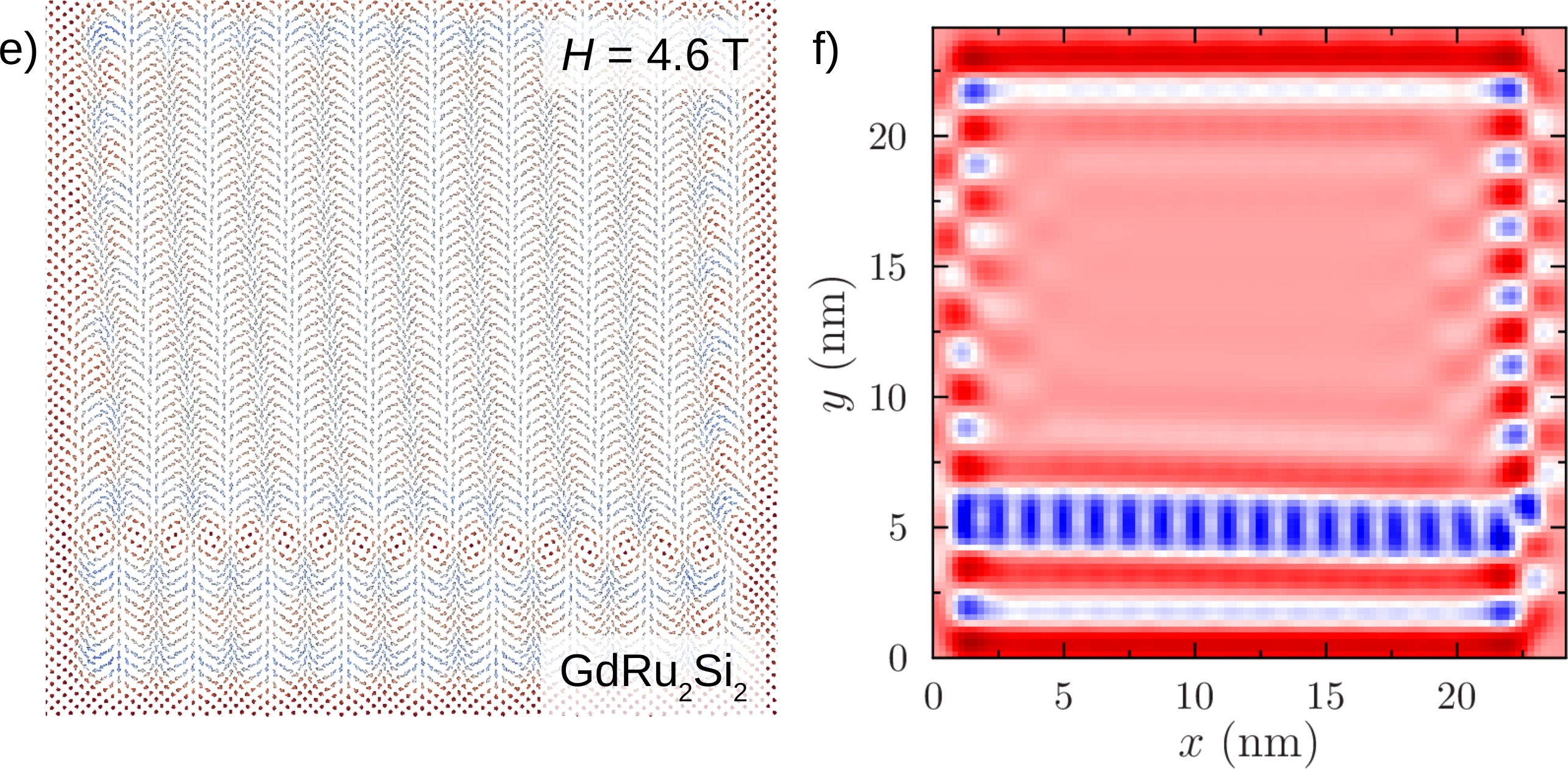}\hspace{10pt}
\includegraphics[width=0.47\textwidth]{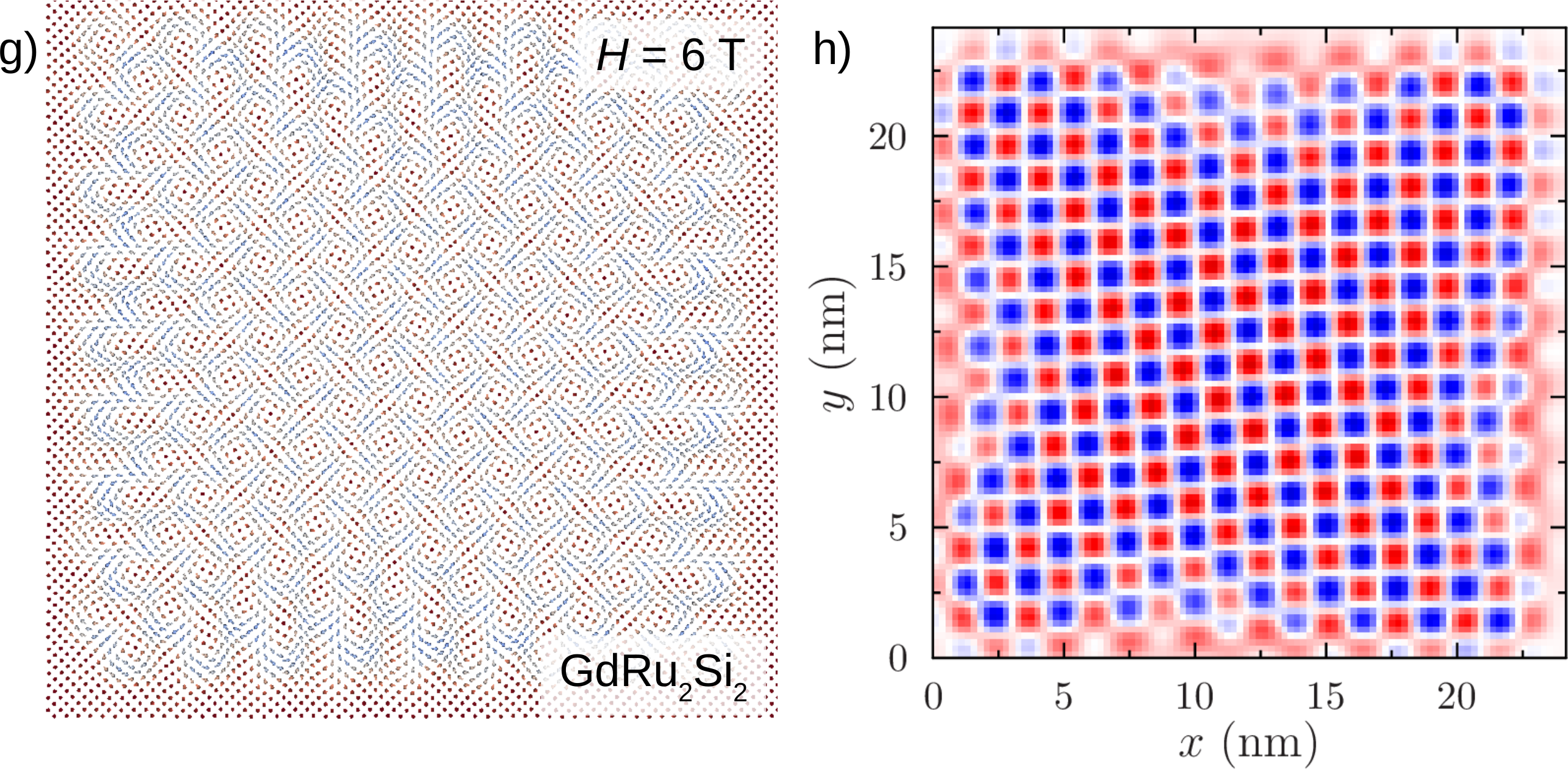}
}
\caption{Real-space distribution of topological charge (spin winding number, Eqn.~(\ref{e:charge})) and corresponding spin configurations in individual Gd layer of GdRu$_2$Si$_2$. Representative cases at different values of external magnetic field are shown. The color code corresponds to the local topological charge from Eqn.~\ref{e:charge}
(red~-- positive value, blue~-- negative value).}
\vspace{-10pt}
\label{f:topological_charge_GdRu2Si2}
\end{figure*}

\begin{figure*}
{\centering
\includegraphics[width=0.47\textwidth]{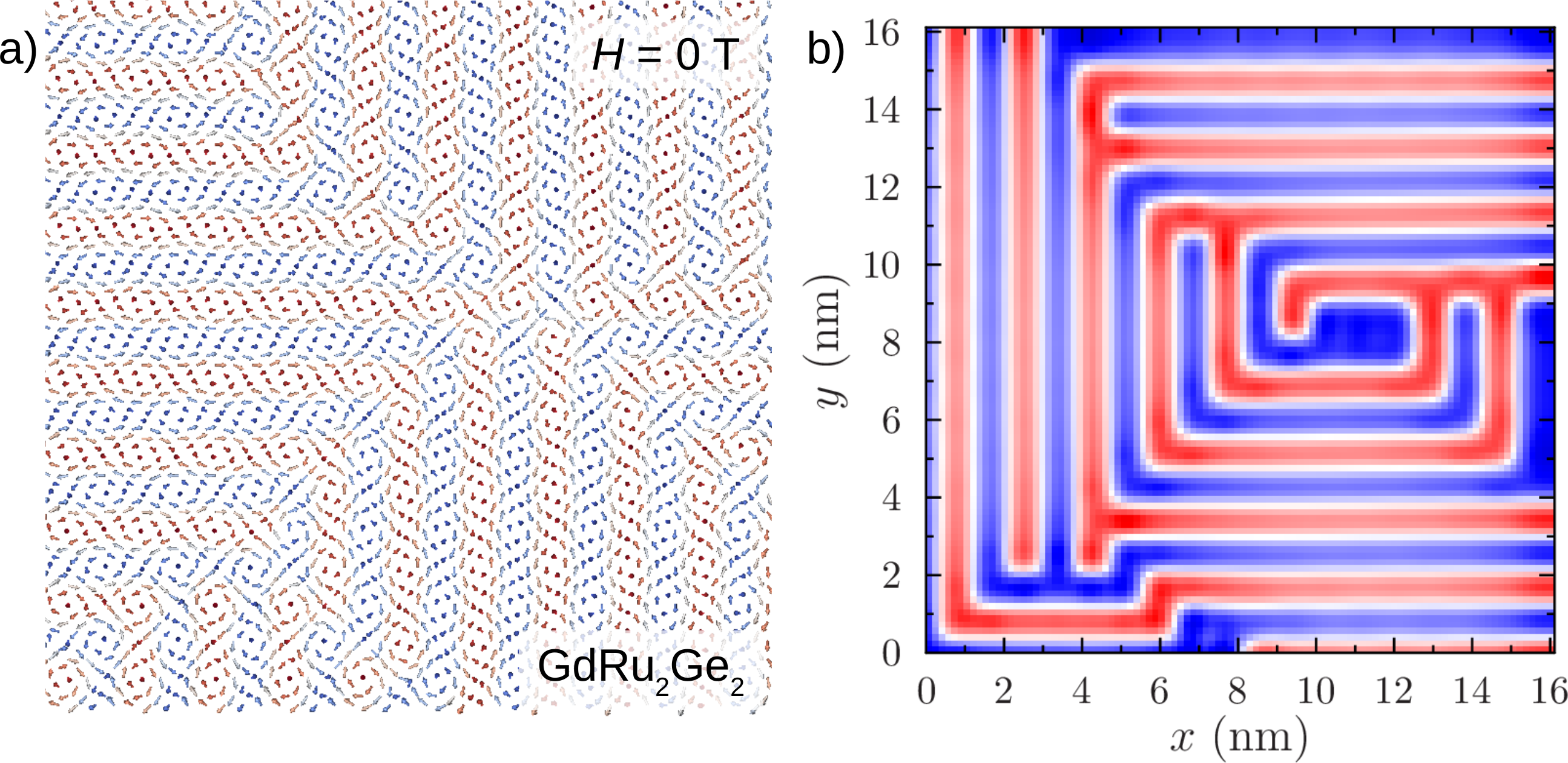}\hspace{10pt}
\includegraphics[width=0.47\textwidth]{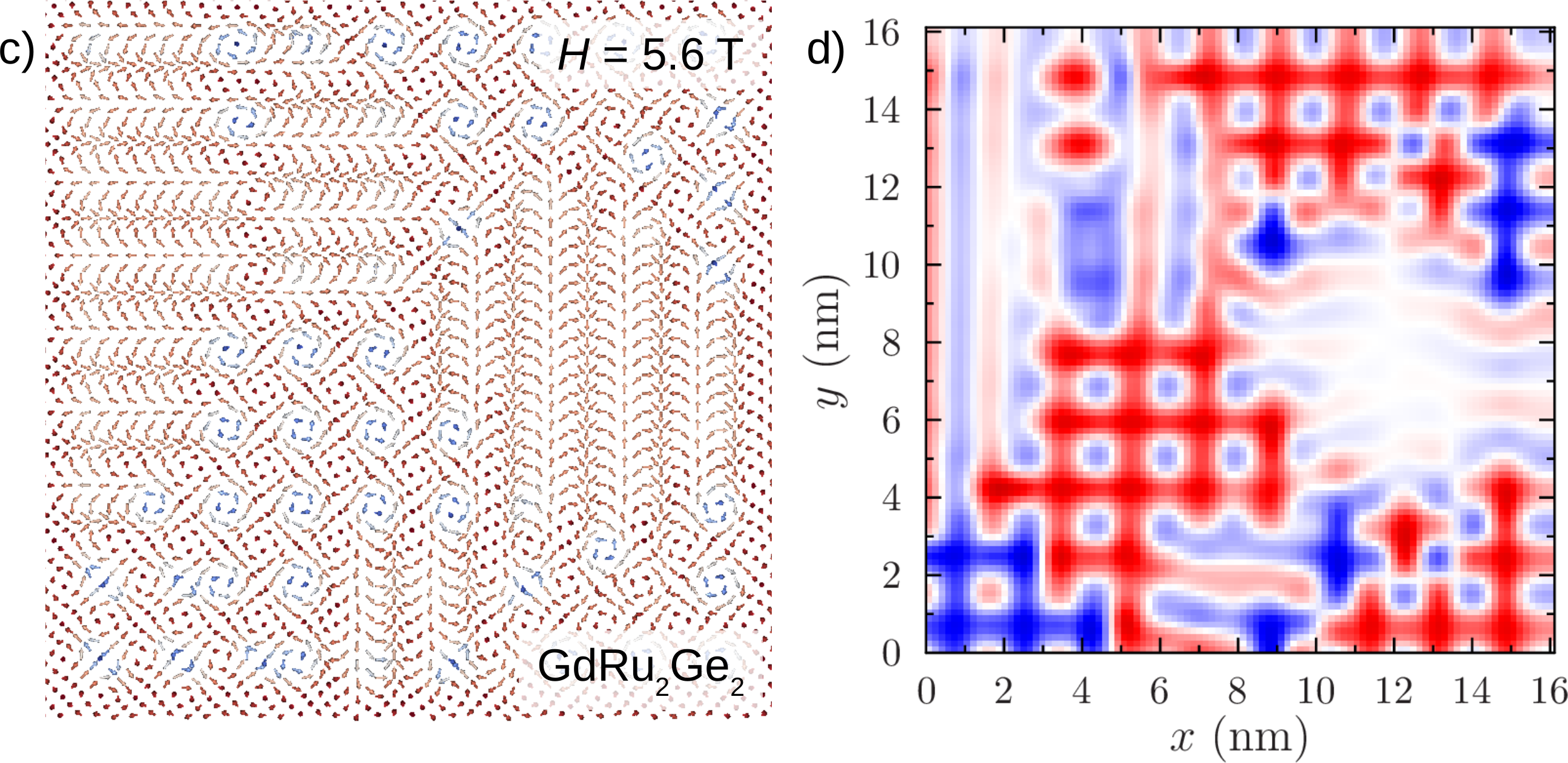}\\[10pt]
\includegraphics[width=0.47\textwidth]{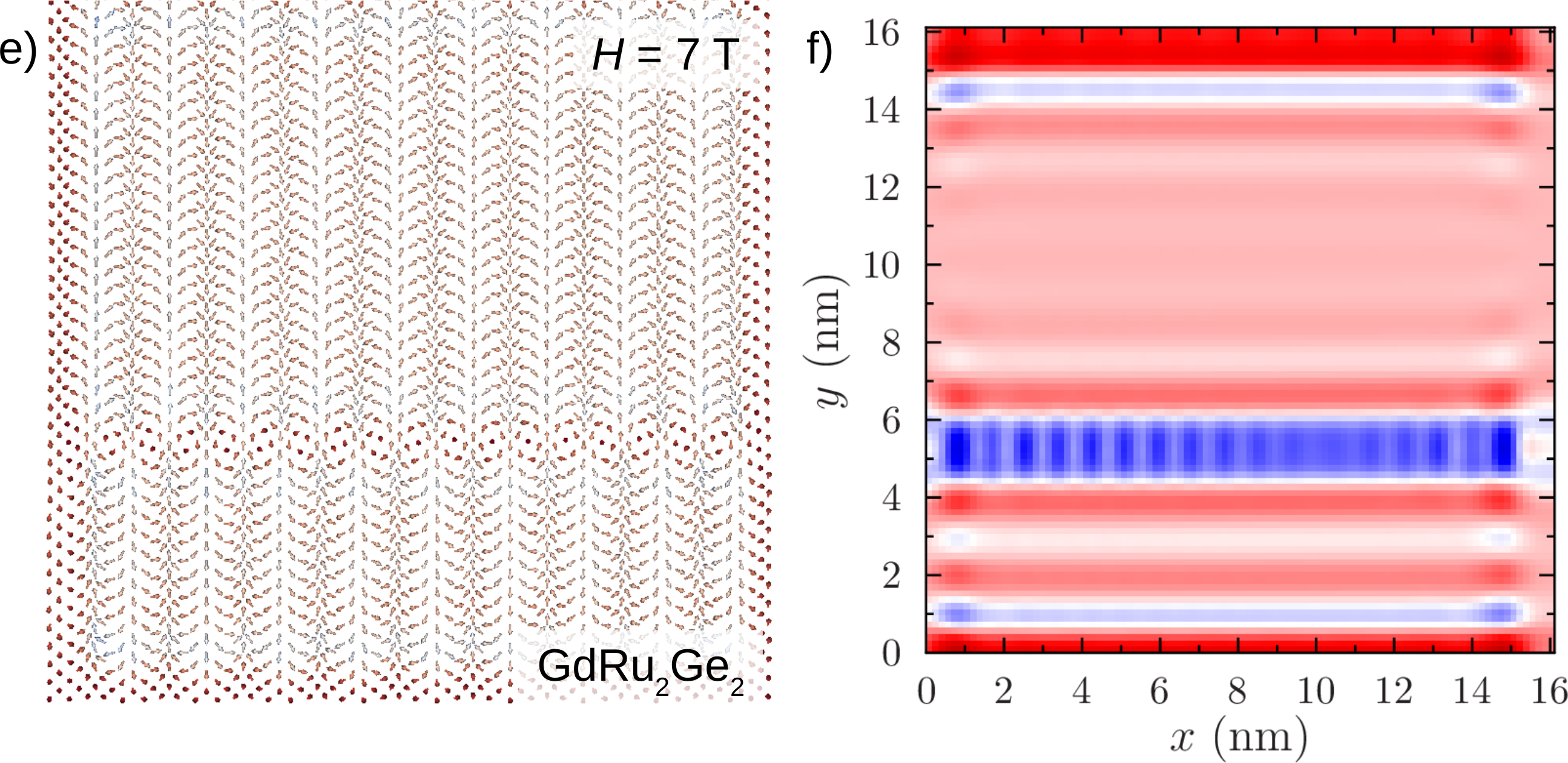}\hspace{10pt}
\includegraphics[width=0.47\textwidth]{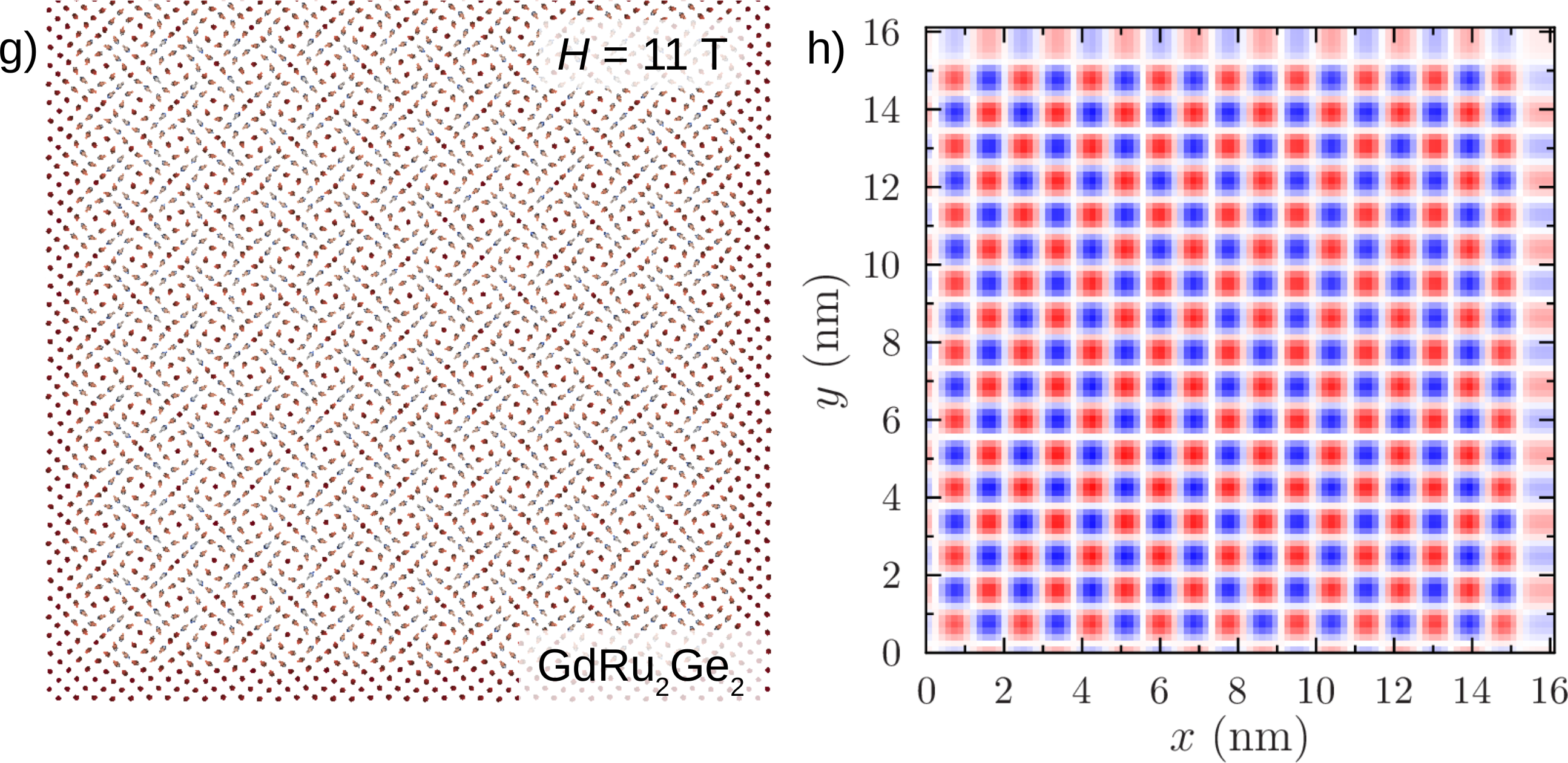}
}
\caption{Real-space distribution of topological charge (spin winding number, Eqn.~(\ref{e:charge})) and corresponding spin configurations in individual Gd layer of GdRu$_2$Ge$_2$. Representative cases at different values of external magnetic field are shown. The color code corresponds to the local topological charge from Eqn.~\ref{e:charge} (red~-- positive value, blue~-- negative value).}
\vspace{-10pt}
\label{f:topological_charge_GdRu2Ge2}
\end{figure*}

\begin{figure*}
{\centering
\includegraphics[width=0.47\textwidth]{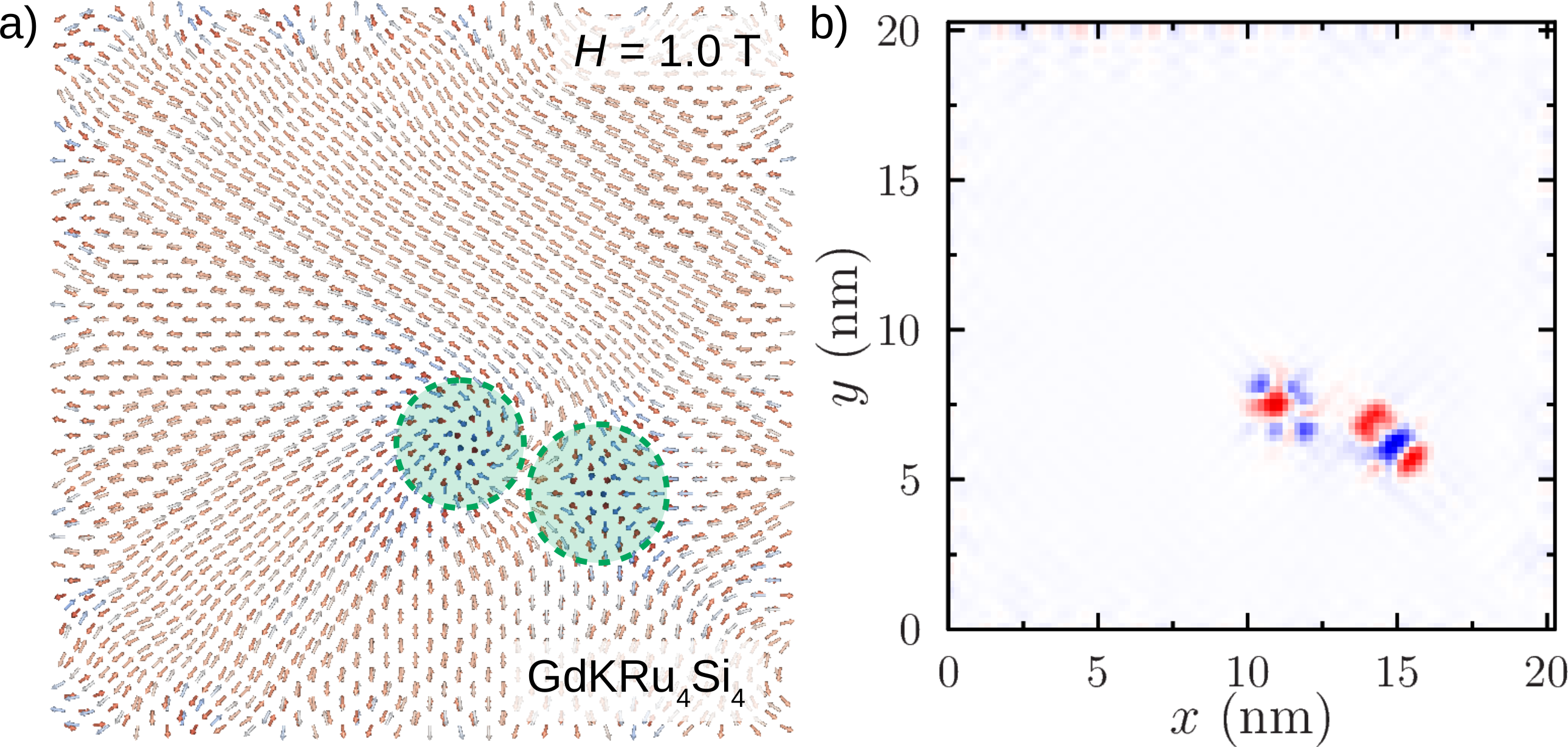}\hspace{10pt}
\includegraphics[width=0.47\textwidth]{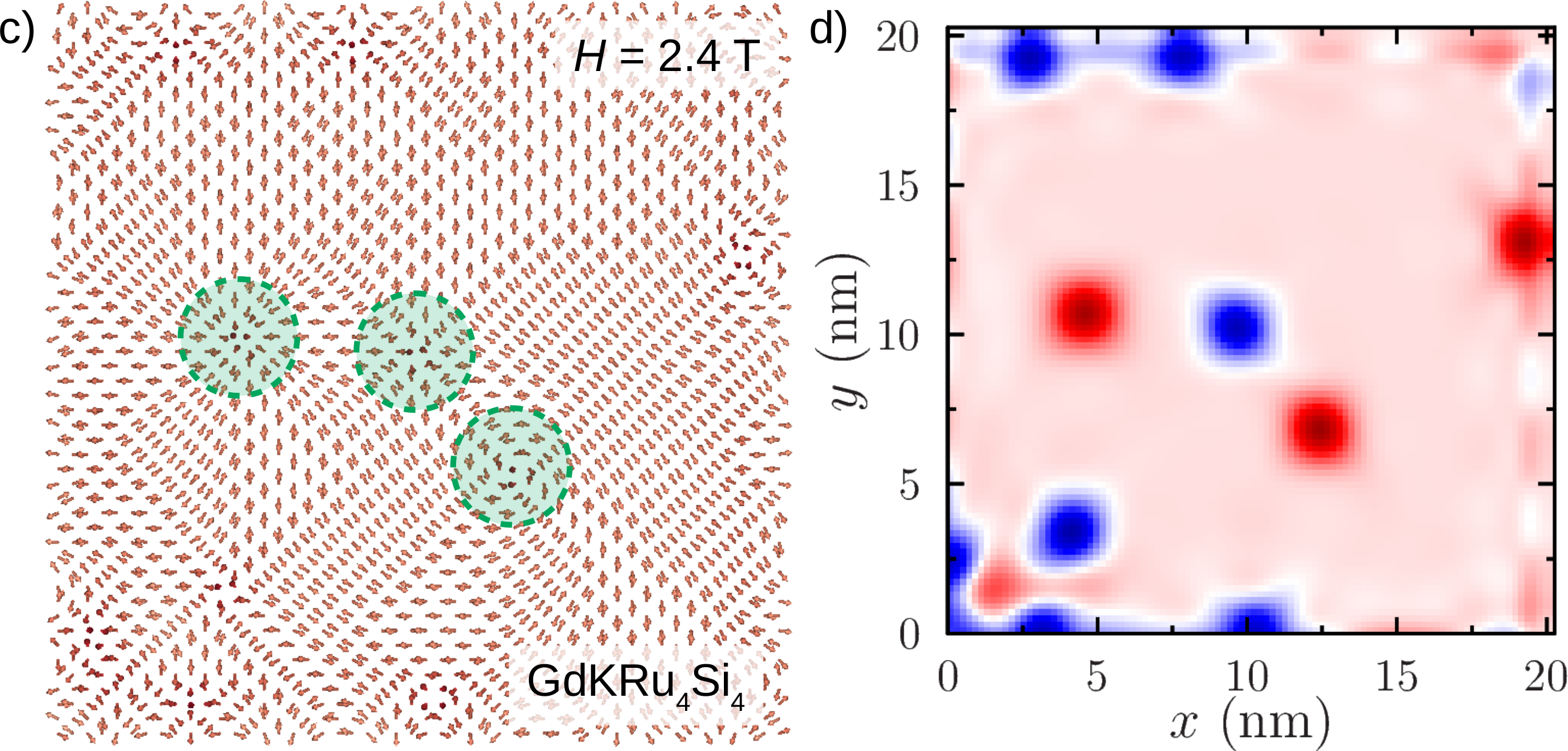}\\[10pt]
\includegraphics[width=0.47\textwidth]{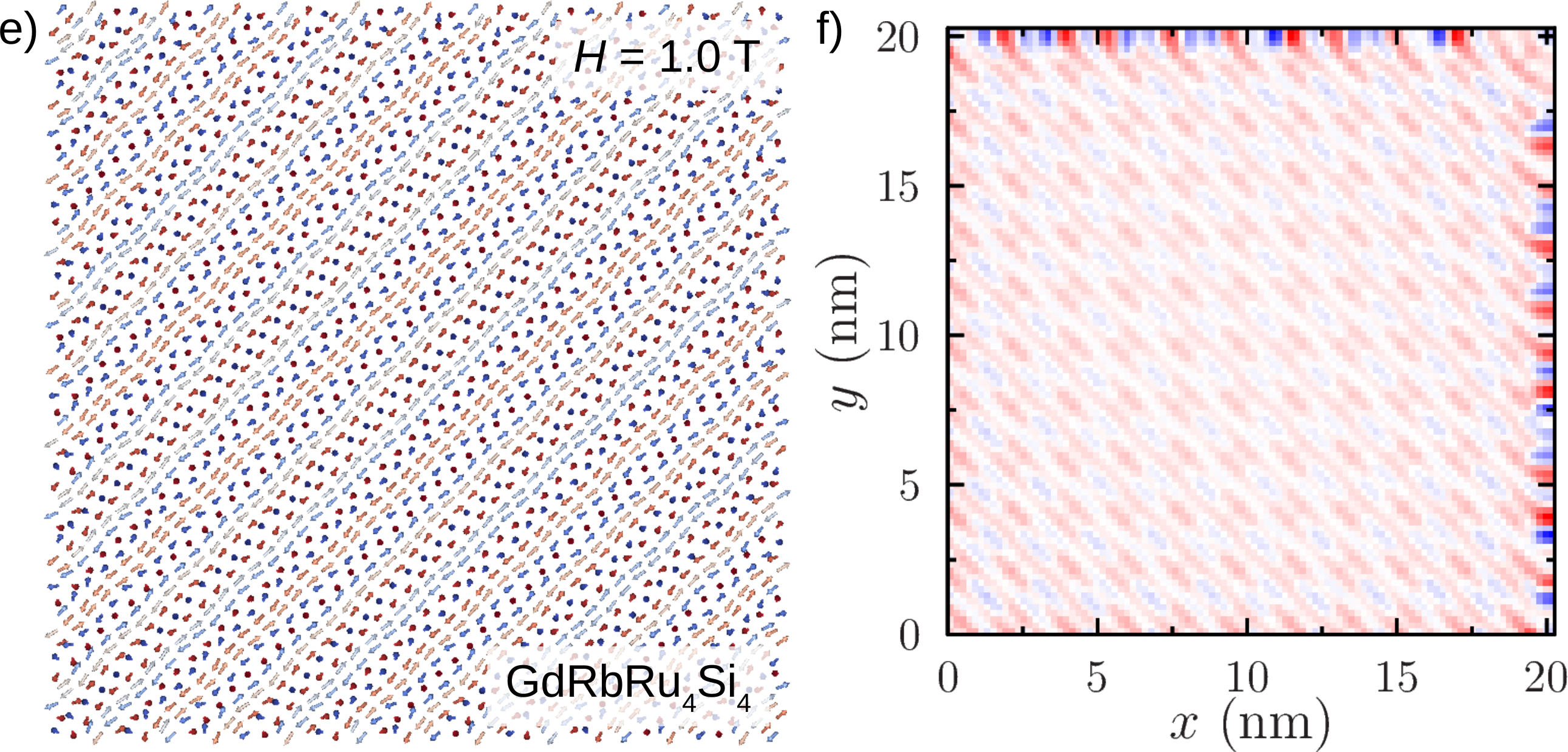}\hspace{10pt}
\includegraphics[width=0.47\textwidth]{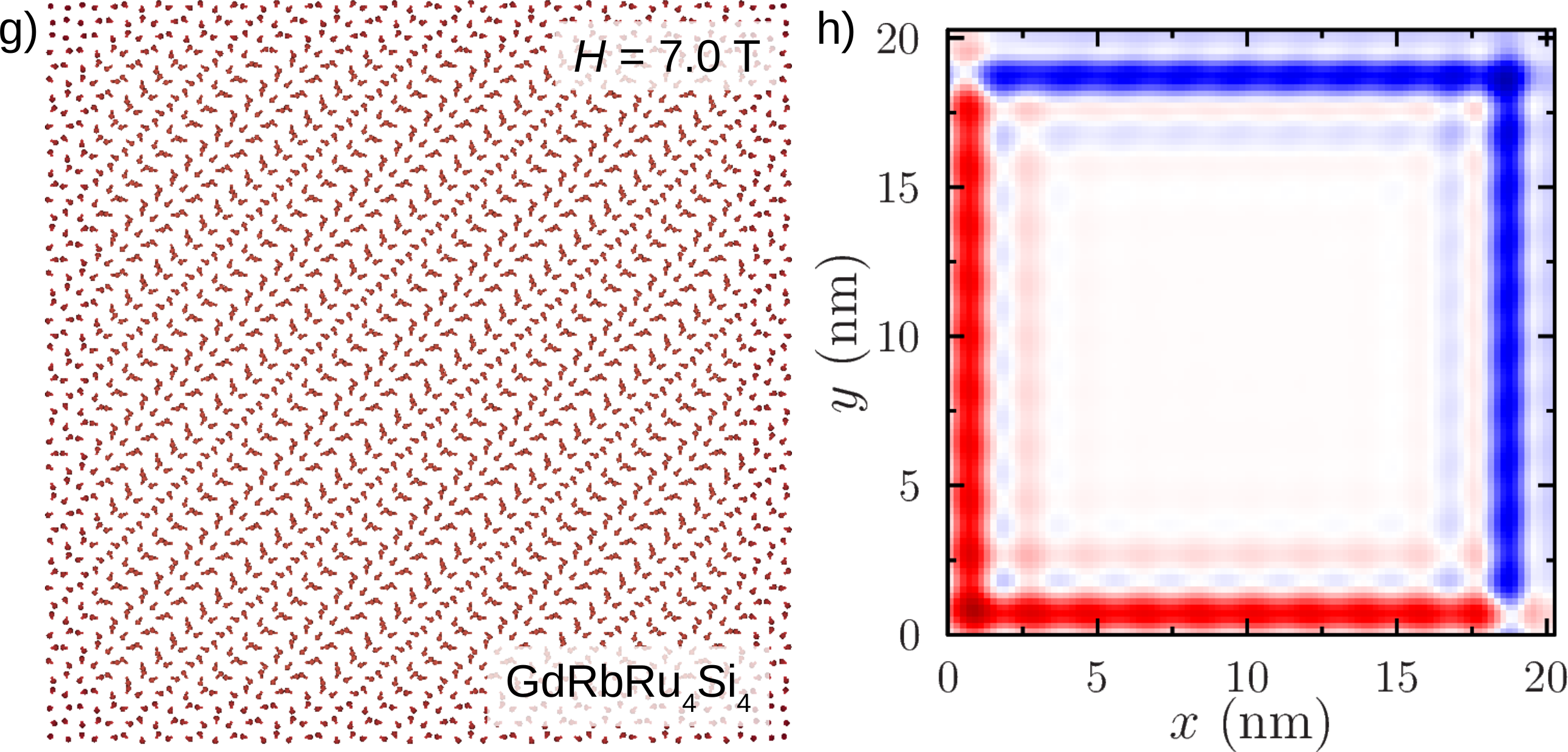}\\[10pt]
\includegraphics[width=0.47\textwidth]{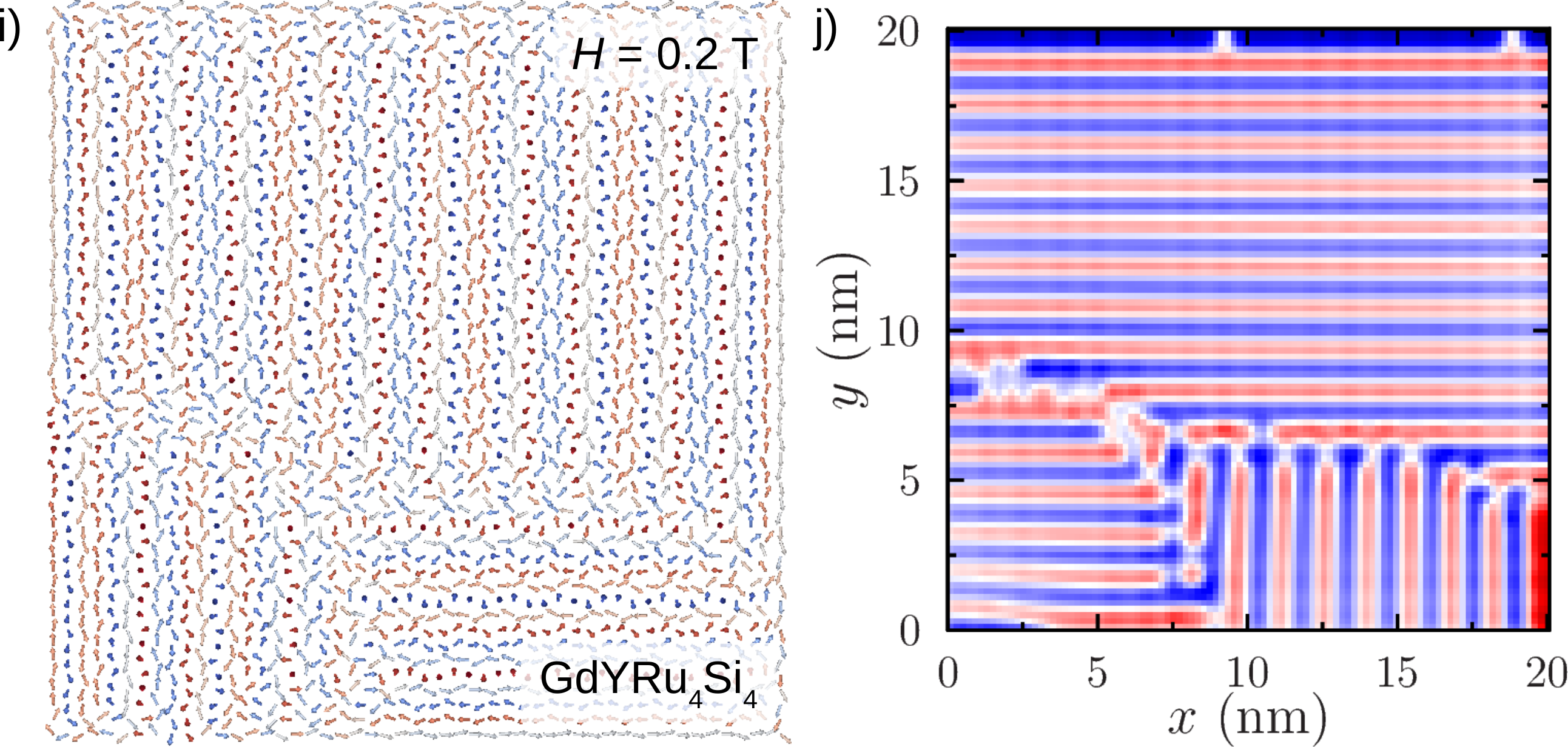}\hspace{10pt}
\includegraphics[width=0.47\textwidth]{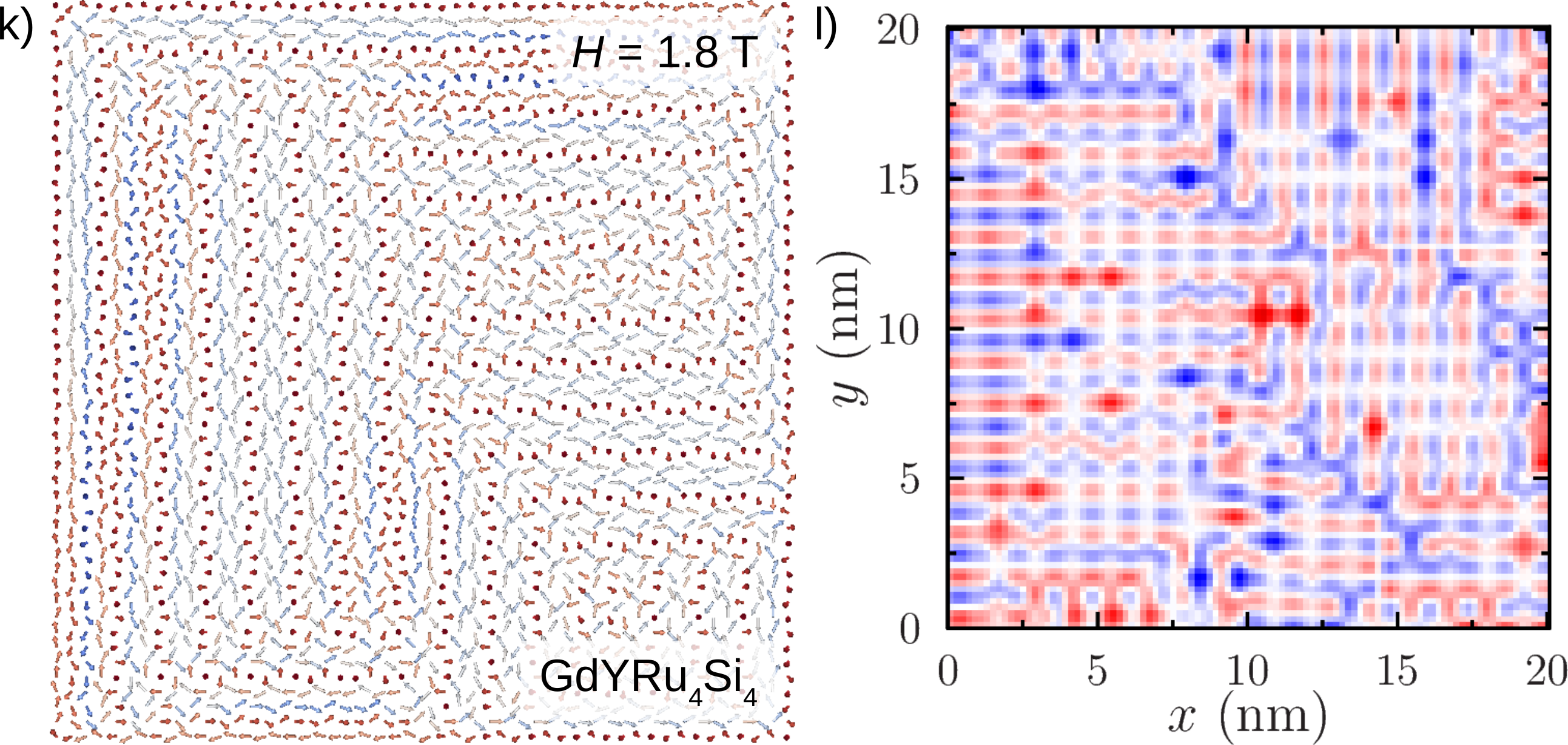}\\[10pt]
\includegraphics[width=0.47\textwidth]{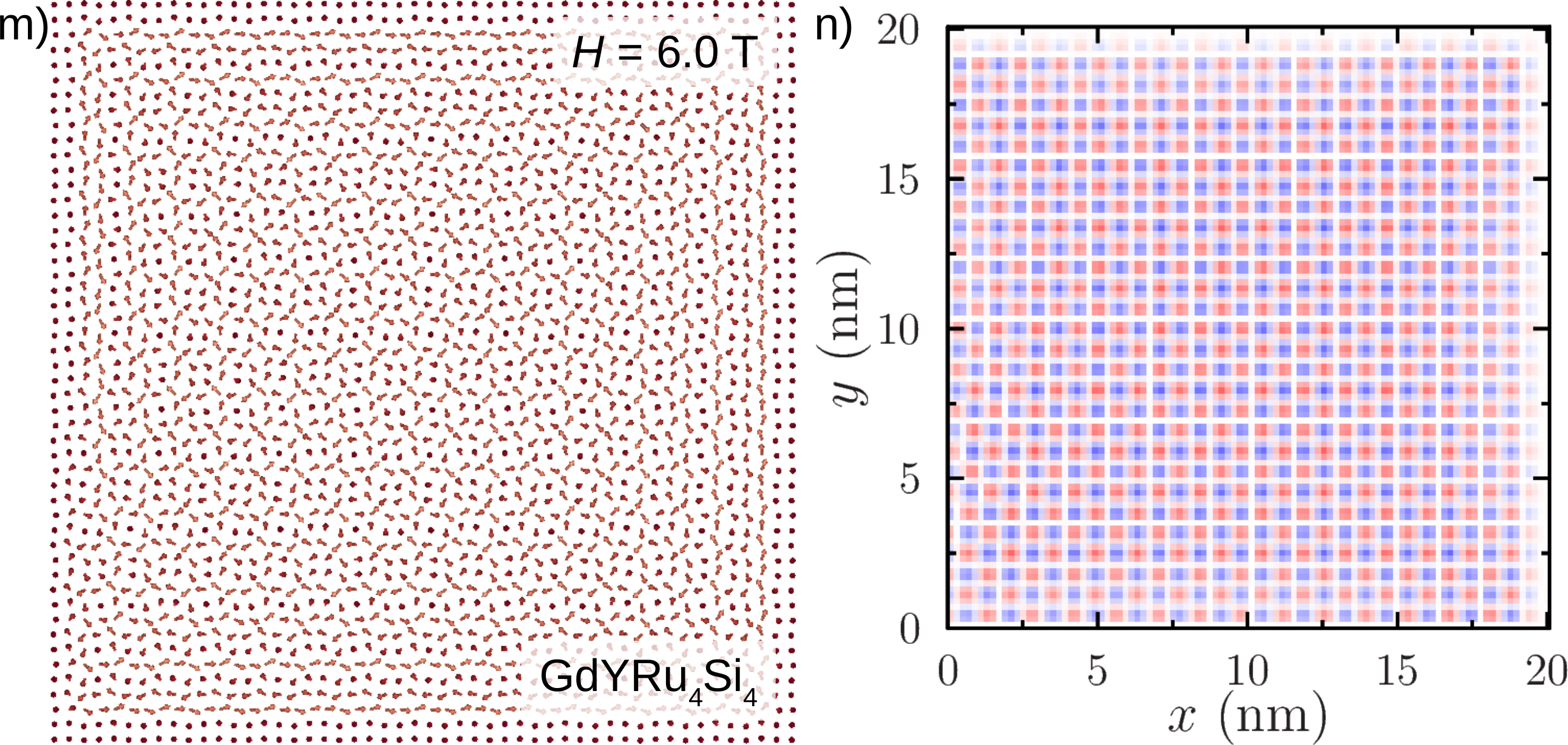}
}
\caption{Real-space distribution of topological charge (spin winding number, Eqn.~\ref{e:charge}) and corresponding spin configurations in individual Gd layer for the 1144-type Gd\textit{A}Ru$_4$Si$_4$ compounds (\textit{A} = K, Rb, Y). Representative cases at different values of external magnetic field are shown. The color code corresponds to the local topological charge from Eqn.~\ref{e:charge} (red~-- positive value, blue~-- negative value).}
\vspace{-10pt}
\label{f:topological_charge_1144}
\end{figure*}

In this section, the real-space distribution of topological charge is presented for the studied systems together with the spin configuration in one of the Gd layers (Figs.~\ref{f:topological_charge_GdRu2Si2}--\ref{f:topological_charge_1144}). The topological charge is calculated using the lattice formulation of the spin winding number, as explained in detail in Ref.~[\onlinecite{Kim2020}]. In general, this quantity helps to distinguish more easily the single-$\vec{Q}$ or double-$\vec{Q}$ character of the magnetic ground state as well as to identify different localized spin textures reminiscent of skyrmions.

\end{document}